%% file: main.tex
\documentclass{article}
\usepackage[utf8]{inputenc}
\usepackage{amsmath}
\usepackage{commath}
\usepackage{gensymb}
\usepackage{url}
\usepackage{cite}
\usepackage{color}
\usepackage{multirow}
\usepackage[margin=1.25in]{geometry}
\usepackage{graphicx}
\graphicspath{{./graphics/}}
\usepackage[font=small,margin=1cm]{caption}
\usepackage{pdfpages}

\newcommand{\ee}{\ifmmode (e,e') \else $(e,e')$\fi}
\newcommand{\Aee}{\ifmmode A(e,e') \else $A(e,e')$\fi}
\newcommand{\eep}{\ifmmode (e,e'p) \else $(e,e'p)$\fi}
\newcommand{\Aeep}{\ifmmode A(e,e'p) \else $A(e,e'p)$\fi}
\newcommand{\eepp}{\ifmmode (e,e'pp) \else $(e,e'pp)$\fi}
\newcommand{\eepN}{\ifmmode (e,e'pN) \else $(e,e'pN)$\fi}
\newcommand{\eepn}{\ifmmode (e,e'pn) \else $(e,e'pn)$\fi}
\newcommand{\pmiss}{\ifmmode p_{miss} \else $p_{miss}$\fi}
\newcommand{\emiss}{\ifmmode E_{miss} \else $E_{miss}$\fi}

\newcommand{\het}{\ifmmode ^3{\rm He} \else $^3$He\fi}
\newcommand{\trit}{\ifmmode ^3{\rm H} \else $^3$H\fi}
\newcommand{\hets}{\ifmmode ^3{\rm He} \else $^3$He \fi} 
\newcommand{\trits}{\ifmmode ^3{\rm H} \else $^3$H  \fi} 

\usepackage{dcolumn}
\newcolumntype{C}[1]{>{\centering\arraybackslash}p{#1}}

\title{Precision measurements of $A=3$ nuclei in Hall B}

\author{
A. Denniston, O.~Hen~(spokesperson), J.~Kahlbow, D.~Nguyen~(spokesperson),\\ 
J.~R.~Pybus, E.P.~Segarra, \\
Massachusetts Institute of Technology, Cambridge, Massachusetts 02139, USA\\
\and
W.~Briscoe, T.~Kutz, S.~Ratliff, A.~Schmidt~(spokesperson), P.~Sharp, I.~Strakovsky\\
The George Washington University, Washington DC 20052, USA\\
\and
C.~Fogler, F.~Hauenstein, C.~Hyde, L.B.~Weinstein~(spokesperson)\\
Old Dominion University, Norfolk, Virginia 23529, USA\\
\and
G.~Johansson, C.~Neuburger, E.~Piasetzky~(spokesperson)\\
Tel-Aviv University, Tel Aviv 69978, Israel\\
\and
D.W.~Higinbotham, C.~Keith, C.~Keppel, J.~Maxwell, D. Meekins~(spokesperson),\\ 
S.~Stepanyan, H.~Szumila-Vance~(contact person)\\
Thomas Jefferson National Accelerator Facility,\\ Newport News, Virginia 23606, USA\\
\and
R.~Cruz Torres\\
Lawrence Berkeley National Laboratory, Berkeley, California 94720, USA\\
\and
A.~Boyer, F.~Benmokhtar, A.~Gadsby, A.~Parker\\
Duquesne University, Pittsburgh, PA 152028 USA\\
\and
L.~Baashen, B.~Raue\\ 
Florida International University, Miami, FL 33199 USA\\
\and
B. McKinnon\\
University of Glasgow, Glasgow, United Kingdom\\
\and
W. Brooks\\
Universidad Técnica Federico Santa María, Valparaíso, Chile\\
\and
A.~Beck, S.~Beck, I.~Korover\\
Nuclear Research Center Negev, Be’er Sheva, Israel\\
\and
Sheren Alsalmi\\
King Saud Univeristy, Riyadh, Saudi Arabia\\
\and
M.~Mihovilovi\v{c}, S.~\v{S}irca\\
University of Ljubljana, Jo\v{z}ef Stefan Institute, Ljubljana, Slovenia\\
\and
C.~Ayerbe, T.~Chetry, D.~Dutta, L.~El Fassi\\ 
Mississippi State University, Mississippi State, MS 39762 USA\\
\and
M.~Osipenko\\
Istituto Nazionale di Fisica Nucleare, Sezione di Genova, Genoa, Italy\\
\and
Eric Christy\\
Hampton University, Hampton, VA, 23669, USA\\
\and
N.~Gevorgyan, N.~Dashyan\\
Yerevan Physics Institute, Yerevan, Armenia\\
\and
Peter Monaghan\\
Christopher Newport University, Newport News, VA 23606, USA\\
\and
R.~Capobianco, S.~Diehl, F.-X.~Girod, K.~Joo, A.~Kim, V.~Klimenko, R.~Santos\\
University of Connecticut, Storrs, CT, USA\\
\and
and the CLAS Collaboration\\
\\
}
\date{}
\begin{document}
\maketitle
\newpage
\begin{abstract}

We propose a high-statistics measurement of few body nuclear structure and short range correlations in quasi-elastic scattering at 6.6 GeV from $^2$H, $^3$He and $^3$H targets in Hall B with the CLAS12 detector. 

We will measure absolute cross sections for $(e,e'p)$ and $(e,e'pN)$ quasi-elastic reaction channels up to a missing momentum $\pmiss \approx 1$~GeV/c over a wide range of  $Q^2$ and $x_B$ and construct the isoscalar sum of \trit{} and \het. We will compare \eep{} cross sections to nuclear theory predictions using a wide variety of techniques and $NN$ interactions in order to constrain the $NN$ interaction at short distances. We will measure  $(e,e'pN)$ quasi-elastic reaction cross sections and $(e,e'pN)/\eep$ ratios to understand short range correlated (SRC) $NN$ pairs in the simplest non-trivial system. $^3$H and $^3$He, being mirror nuclei, exploit the maximum available isospin asymmetry.  They are light enough that their ground states are readily calculable, but they already exhibit complex nuclear behavior, including $NN$ SRCs.  We will also measure $d\eep$ in order to help theorists constrain non-quasielastic reaction mechanisms in order to better calculate reactions on $A=3$ nuclei.  Measuring all three few body nuclei together is critical, in order to understand and minimize different reaction effects, such as single charge exchange final state interactions, in order to test ground-state nuclear models.

We will also measure the ratio of inclusive $(e,e')$ quasi-elastic cross sections (integrated over $x_B$) from $^3$He and $^3$H in order to extract the neutron magnetic form factor $G_M^n$  at small and moderate values of $Q^2$.  We will measure this at both 6.6 GeV and 2.2 GeV.

Simulations and previous experience with CLAS12 show that we can obtain meaningful statistics with 55 days of beam time. We request 20 beam days each on $^3$He and $^3$H, and 10 beam days on $^2$H at an incident beam energy of 6.6 GeV. We request an additional 5.5 days of beam time at 2.2 GeV  to measure $G_M^n$ at low $Q^2$.  Our proposed measurements will significantly constrain nuclear models for few body nuclear dynamics and short range structure. Also, we will measure the behavior of the neutron magnetic form factor at low to moderate $Q^2$.  

This experiment will use the base equipment for CLAS12 and requires the construction of a new $^3$H target.  The JLab Target Group has a preliminary target design. The groups working on this experiment bring previous experience working with a $^3$H target in Hall A. 
\end{abstract}

\section{Introduction}\label{intro}

We propose to measure quasi-elastic electron scattering on $^3$He, $^3$H, and deuterium targets in Hall B using the CLAS12 detector in its standard configuration (no forward tagger) with an open electron trigger. We will use a newly designed cryo-target that is specifically developed for the $^3$H containment and safety considerations (designed and developed with previous experience from the Hall A $^3$H measurements, see section~\ref{TargetSection}). We will use identical cryo-target cells for the \trit, \het, and $^2$H targets. The CLAS12 detector can access a large range in $x_B$ and $Q^2$ allowing us to measure reactions over a wide kinematic coverage. This will let us select kinematics where the effects of outgoing nucleon rescattering (final state interactions (FSI)) and other non-quasi-elastic reaction mechanisms are minimized.

The experiment  will  measure \hets and \trit\eep{} proton momentum distributions to extremely high missing momenta ($\vec p_{miss}= \vec p\thinspace' - \vec q$, where $\vec q$ is the three-momentum transfer and $\vec p\thinspace'$ is the outgoing proton momentum) in order to constrain the short distance behavior of models of the $NN$ potential. It will also measure the other nucleons knocked out in the reaction using $(e,e'pp)$ and $(e,e'pn)$ to measure characteristics of nucleon-nucleon ($NN$) Short Range Correlations (SRC).  These $A=3$ mirror nuclei provide the maximum isospin asymmetry, coupled with theoretical calculability and minimal attenuation or rescattering of the outgoing hadrons. We will also measure $d\eep$ to constrain calculations of non-quasielastic reaction mechanisms on the simplest nucleus and to determine kinematics least sensitive to these effects. The combination of \het{} and \trit{} is essential as theoretical predictions must be able to account for both. As will be discussed in Section~\ref{overview}, measurements of each individual $A=3$ nuclei alone are insufficient without the other in order to fully constrain the $NN$ interaction and decouple the contributions of FSIs.

In addition, we will measure quasi-elastic inclusive scattering  on \hets and \trits to extract the neutron magnetic form factor $G_M^n$ at low to moderate $Q^2$. A short 2.2-GeV run will enable us  to measure $G_M^n$ at  $Q^2<0.5$~[GeV/$c]^2$ where previous measurements are in contention with model predictions. 

We build on the Letter Of Intent~\cite{loiA3} that describes experimental possibilities with a $^3$H target in Hall B, but we focus on  quasi-elastic scattering. The CLAS12 detector is ideally-suited for these studies with only slightly lower luminosity than the previous Hall A measurements but significantly larger acceptance. The combination of luminosity and acceptance enable high statistics studies with access to a wide range of kinematical dependencies that will be able to precisely guide nuclear theory. 

We present an overview of the current physics that is relevant to the measurements we will obtain in $A=3$ nuclei  in Section~\ref{overview}. We detail the physics goals in Section~\ref{physGoals}, and we discuss the measurements and relevant observables of the quasi-elastic reaction in detail in Section~\ref{QESection}. Finally, we provide a description and plan for the design of the $^3$H target (Section~\ref{TargetSection}).

\section{Overview of recent results}\label{overview}

The fundamental dynamics of the nuclear many-body system has implications in other areas of physics from understanding astrophysical systems such as neutron stars to discerning the many body fermion systems in cold atomic physics. The nuclear few-body system allows us to test critical aspects of our understanding of many-body systems, specifically the $NN$ interaction at short distances and the effects of short range correlated nucleon pairs.  Nuclear few-body systems are relatively simple, calculable systems and measurements of these systems can provide precise tests of modern theories.

Quasi-elastic (QE) electron scattering is sensitive to the ground state properties of nuclei. For heavy nuclei, this sensitivity is difficult to evaluate due to imprecise nuclear ground state calculations and contributions from non-QE reactions. The non-QE contributions have significant and hard-to-quantify impacts on the measured cross sections. These contributions including final state interactions (FSIs), single charge exchange (SCX), meson exchange currents (MEC), and $\Delta$ production or isobar configurations (IC), are highly model-dependent and obscure the interpretation of data in terms of the nuclear ground state~\cite{Ford_2014}.

Studies of both \het{} and \trit{} resolve the complications that arise when calculating heavier nuclei. They have exactly calculable ground states from nuclear-interaction models. Furthermore, the combination of measurements from \het, \trit, and deuterium over a larger kinematic range can be used to uniquely disentangle the contributions from FSIs, optimize the kinematic selection for this reduction and thus enable one to directly relate measured cross-sections to the ground-state momentum distributions. Early studies on \hets alone showed that we are sensitive to the ground state distributions and can make detailed comparisons with theory~\cite{Marchand}. Here, the combination of studying \het{} and \trit{} nuclei benchmark modern nuclear theory and place tight constraints. 

Deuterium, \het{}, and \trit{} put to test the most fundamental descriptions of $eA$ reactions, including how an electron interacts with an off-shell bound nucleon. Recent theoretical work~\cite{Vera_2018} using light front calculations (i.e. light cone reference frame) provides techniques to reduce off-shell effects in the electron-bound-nucleon cross section at $Q^2>1$ GeV$^2$/c$^2$ to much less than in the standardly-used deForest prescriptions~\cite{DeForest:1983ahx}. This effect is separate from bound nucleon modifications. We plan to work closely with theorists to better understand and minimize this effect. 

$NN$ SRC pairs have remarkably similar  behavior in all nuclei~\cite{Hen2016kwk} and understanding them is critical for understanding the short-distance and high-momentum behavior of nucleons. SRCs are  pairs of nucleons with  relative momenta $p_{rel}$ much greater than typical mean-field nucleon momenta  and  center-of-mass $p_{cm}$ momenta consistent with mean-field momenta.  Almost all high-momentum nucleons in nuclei belong to an SRC pair~\cite{Subedi2008zz}.  When nucleons are at short relative distances, they experience a strong short-ranged interaction that generates these high relative momenta. 

Experimental work has shown that about 20\% of nucleons in medium to heavy nuclei belong to SRC pairs and these pairs are predominantly neutron-proton pairs (at relative momenta of 300 to 600 MeV/c)~\cite{Hen2016kwk, Atti2015eda, Carlson2014vla, Arrington2011xs, Frankfurt2008zv, Hen2014nza, Korover2014dma, Subedi2008zz, Piasetzky2006ai, Fomin2011ng, Egiyan2005hs, Frankfurt1993sp}. As nucleons are composite objects,  their internal structure (quark distributions) may be modified when the distance between the nucleons is smaller than their radii. In this way, the study of SRCs can give us information about bound nucleons and nuclear structure.  

The study of SRCs is the subject of much experimental and theoretical work, as well as phenomenological applications to other areas of physics. A detailed overview of SRC physics is found in Ref.~\cite{Hen2016kwk} and a theory-oriented description is in Ref.~\cite{Atti2015eda}. The following discussion will focus on the most recent results with their implications for $A=3$ nuclei. 

\subsection{Few-body nuclear structure}

Light nuclei are  ideal for studying the nuclear system. In particular,  $A=3$ nuclei (mirror nuclei $^3$He and $^3$H) play a unique role in  nuclear structure studies. This system is complex enough to include some fundamental nuclear environment effects but simple enough that its ground state can still be exactly calculable. As will be detailed below, measurements on both \hets and \trits tightly constrain the reaction mechanism and $NN$ interaction. Theory must be able to explain the data on both nuclei, and the combination of measurements on both is necessary to adequately account for the non-QE contributions. Consequently, the absolute cross section measurement on the $A=3$ system can be used as a test for nuclear theory calculations.

The momentum distributions of $A=3$ nuclei uniquely benchmark modern nuclear theory. In the QE limit with no re-scattering effects, the initial momentum of the probed nucleon (in this case, proton) is measured as \pmiss. The simultaneous measurements of both \het{} and \trit{} cross sections tightly constrain the contribution of non-QE reactions to our measurement, thus improving the purity of \pmiss{} as it relates to the proton's initial momentum. By measuring both \het{} and \trit{}, we improve our sensitivity to access the ground states. 

\begin{figure}[htb]
\centering
\includegraphics[width= 0.45\textwidth]{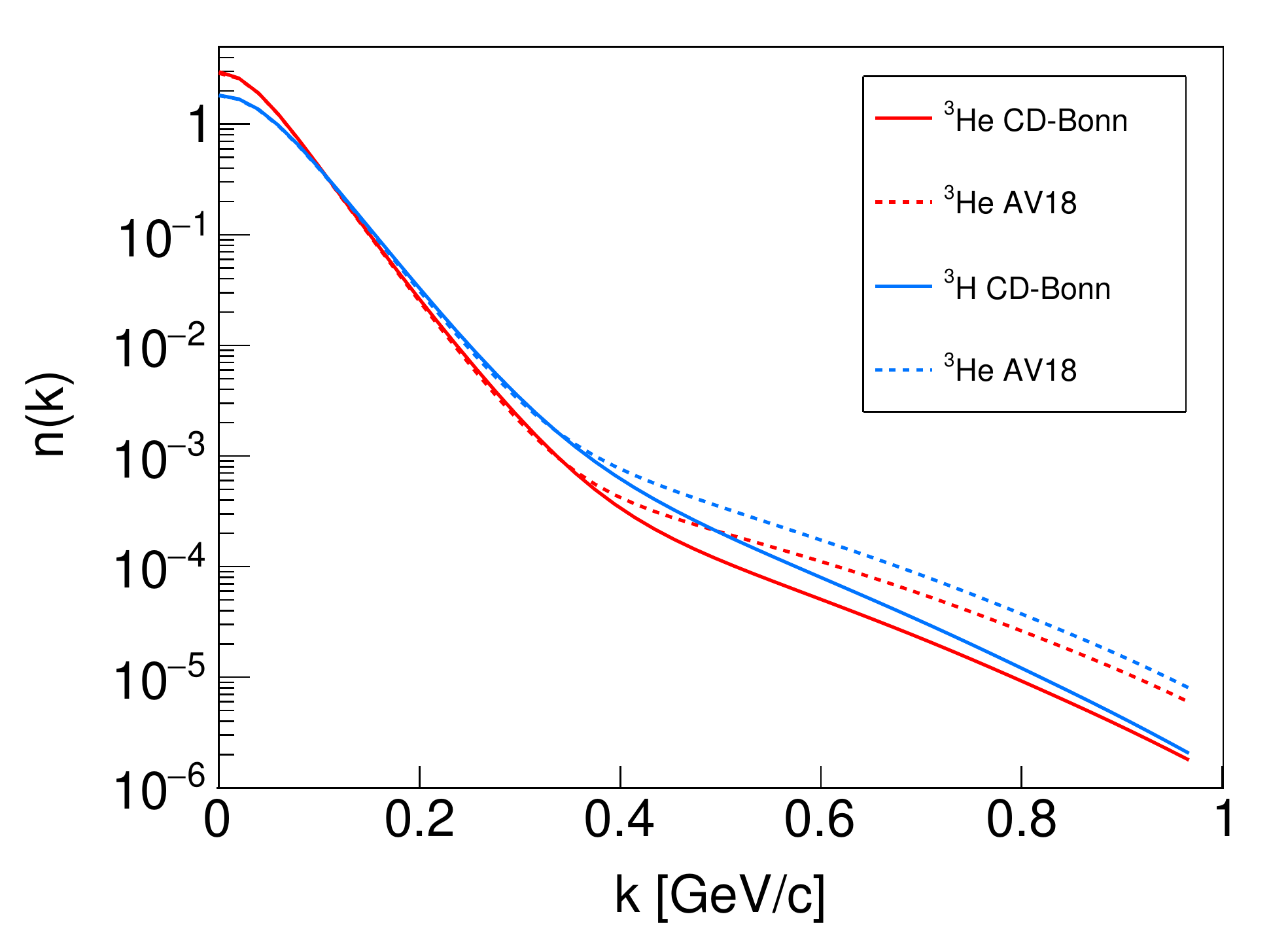}
\includegraphics[width= 0.45\textwidth]{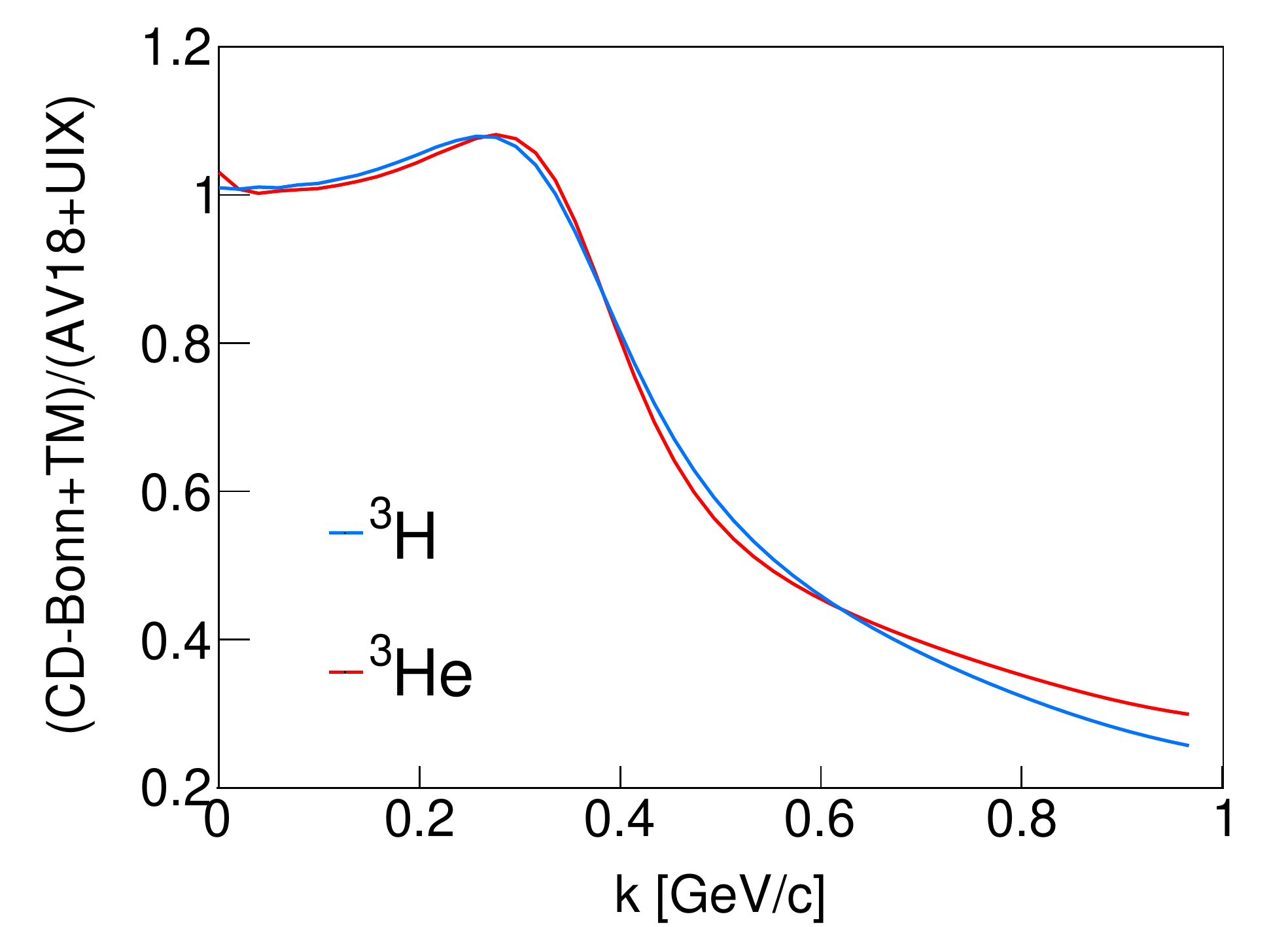}
\caption{Left: The momentum distributions for \het{} and \trit{} are shown for CD-Bonn+TM and AV18+UIX as a function of the proton momentum. Right: The ratio of $^3$He  and $^3$H proton momentum distributions are shown using the CD-Bonn+TM potential relative to the AV18+UIX potential using the calculation in Ref.~\cite{Marcucci:2018llz}.}
\label{fig:mom_distr}
\end{figure}

As in the Hall A tritium measurements, we will measure absolute cross sections of \eep. We use various $NN$ interaction models to predict the momentum distributions. Two primary momentum distributions are those from CDBonn-TMD and AV18+UIX as shown in Fig.~\ref{fig:mom_distr} as a function of the proton momentum. We do not include predictions from Chiral Effective Field Theories as they are not valid for the high proton momentum (and small nucleon separation).

The right side of Fig.~\ref{fig:mom_distr} shows the ratio of the $^3$He and $^3$H proton momentum distributions obtained using the CD-Bonn+TM potential relative to that obtaining using the AV18+UIX potential calculation in Ref.\cite{Marcucci:2018llz}. The different momentum distributions for \hets and \trits agree at low $p$, where the $NN$ potentials are well constrained by $NN$ scattering. At larger $p$, where the $NN$ potentials are not well constrained and where pion degrees of freedom make it harder to calculate $NN$ potentials, these distributions begin to diverge. 

\begin{figure}[htb]
\begin{center}
\includegraphics[width =0.7\textwidth, page=1]{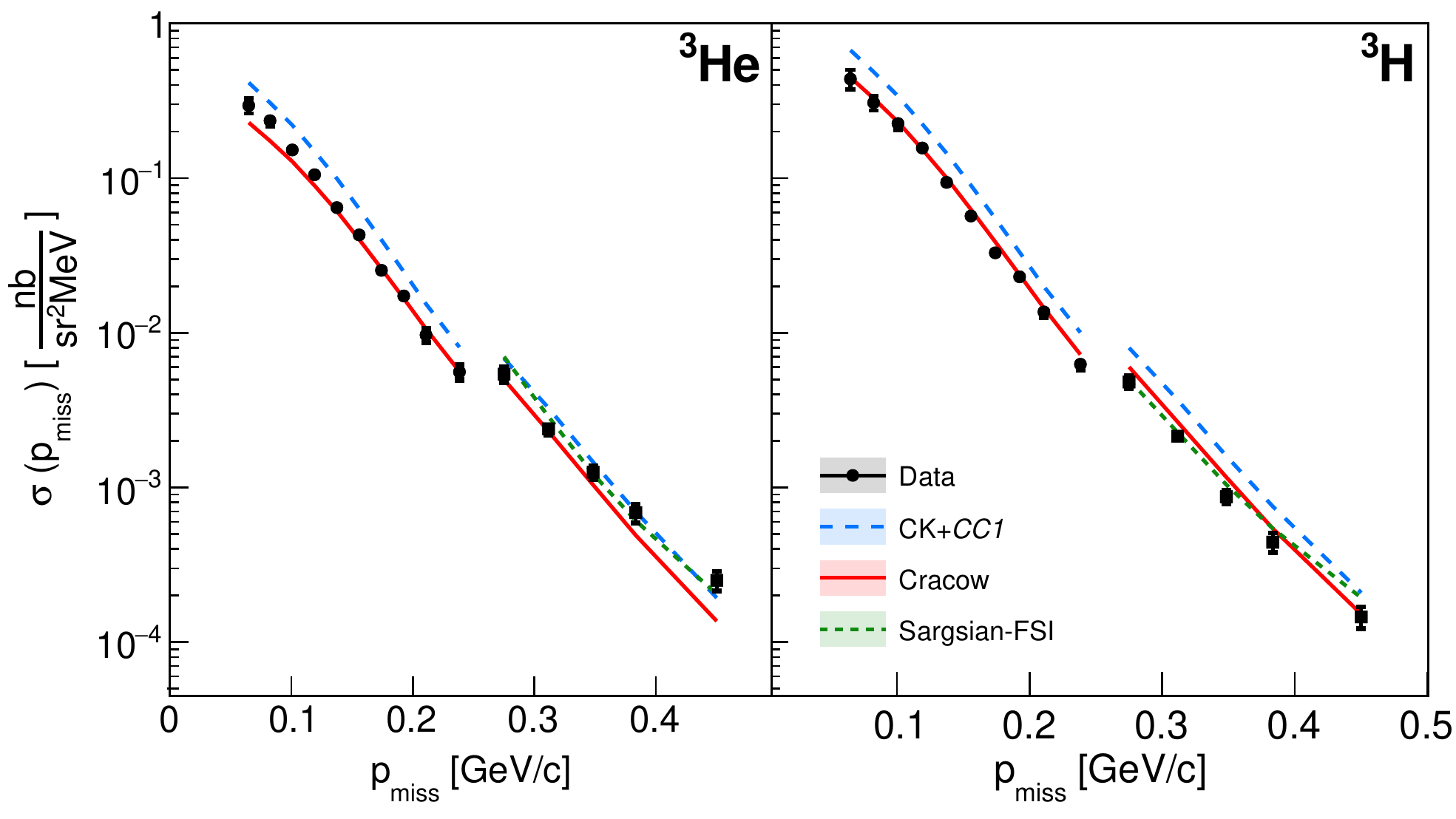}
\end{center}
\caption{Absolute cross section as a function of \pmiss{} for $^3$He (left) and $^3$H (right). The different sets of data points, depicted by black circles and squares, correspond to the cross sections measured in the low-\pmiss{} and high-\pmiss{} kinematical settings, respectively. The lines correspond to cross sections calculated from different theoretical models: Cracow (solid red), CK+$CC1$ (dashed blue) and Sargsian-FSI (dotted  green, $\pmiss>250$ MeV/c only). The different kinematical settings have different average elementary electron-nucleon cross-sections and therefore have a different overall scale for both data and calculations.}
\label{fig:AbsCross}
\end{figure}

The recent Hall A measurements measured the absolute cross section of \trit{} and \het{}~\cite{Cruz-Torres:2019bqw,Cruz-Torres:2020uke} and compared the cross sections with state-of-the-art ab-initio calculations. The results are shown in Fig.~\ref{fig:AbsCross}. 

\begin{figure}[!htb]
\includegraphics[width = 6.5cm, page=2]{Figures/XSection_A3.pdf}
\includegraphics[width = 6.5cm, page=1]{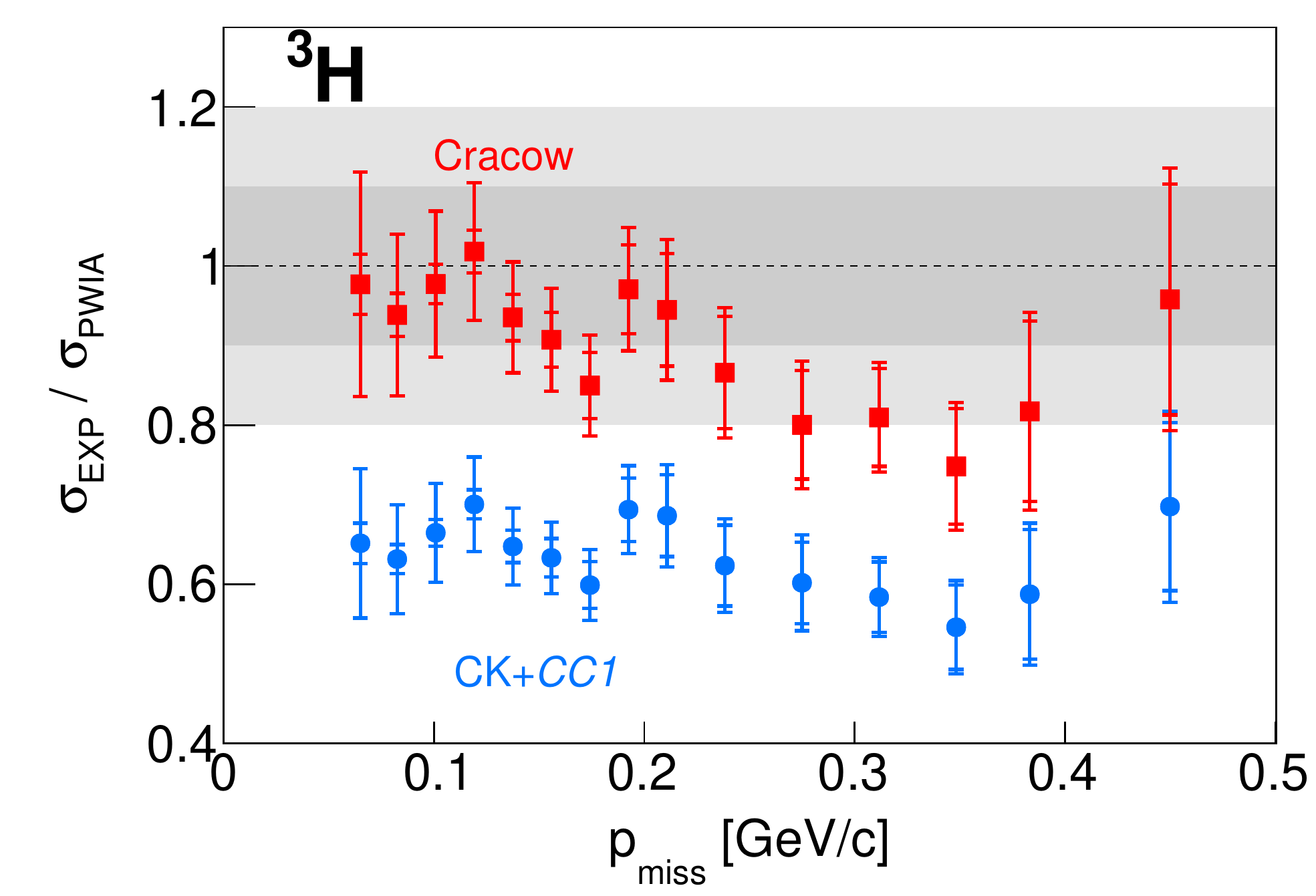}
\caption{The ratio of the experimental cross section to different PWIA calculations plotted versus $p_{miss}$ for $^3$He\eep{} (left) and $^3$H\eep{} (right). Red squares show the ratio to the Cracow calculation while blue circles show the ratio to the Ciofi-Kaptari spectral-function-based calculations (CK+$CC1$). Open symbols show the $^3$He\eep{} data of Ref.~\cite{Benmokhtar:2004fs}, taken at lower $Q^2$ and $x\sim1$ kinematics, compared with the PWIA calculations of Ref.~\cite{CiofidegliAtti:2005qt,Laget:2004sm,Frankfurt:2008zv,Alvioli:2009zy}. The inner and outer bars show the statistical and statistical plus systematic uncertainties, respectively. The shaded regions show $10\%$ and $20\%$ agreement intervals~\cite{Cruz-Torres:2020uke}.}
\label{fig:resultsPWIA}
\end{figure}

These cross section measurements were taken at large momentum transfer ($\langle Q^2 \rangle \approx 1.9$ (GeV/$c)^2$) and $x_B>1$ kinematics, which minimizes contributions from MECs and ICs. A further requirement on the angle between momentum transfer and the missing momentum, $\theta_{\pmiss{}q}<40^\circ$, reduced the effects of FSIs.  Thus, the measured cross sections were relatively sensitive to QE scattering from single nucleons. The data covered missing momenta $40 \le \pmiss \le 500$~MeV/c. The ratio of the measured to calculated cross sections is shown in Fig.~\ref{fig:resultsPWIA}~\cite{Cruz-Torres:2020uke}.  

The data and ab initio PWIA calculations by the Cracow Group~\cite{Golak2005ElectronAP, caracsco} (which include the continuum interaction between the two unstruck nucleons, but not the rescattering of the struck nucleon) agreed to within about 20\% for $^{3}$H for  the full \pmiss{} range. The difference between data and calculation was within about 20\% also for $^3$He up to \pmiss{} of 350~MeV/c.  This validates the choice of kinematics and shows that the Hall A measurements have significantly reduced the effects of nucleon re-scattering, so that the measured cross sections are sensitive to the underlying ground state distributions.

\begin{figure}[htb]
\begin{centering}
\includegraphics[width =\textwidth]{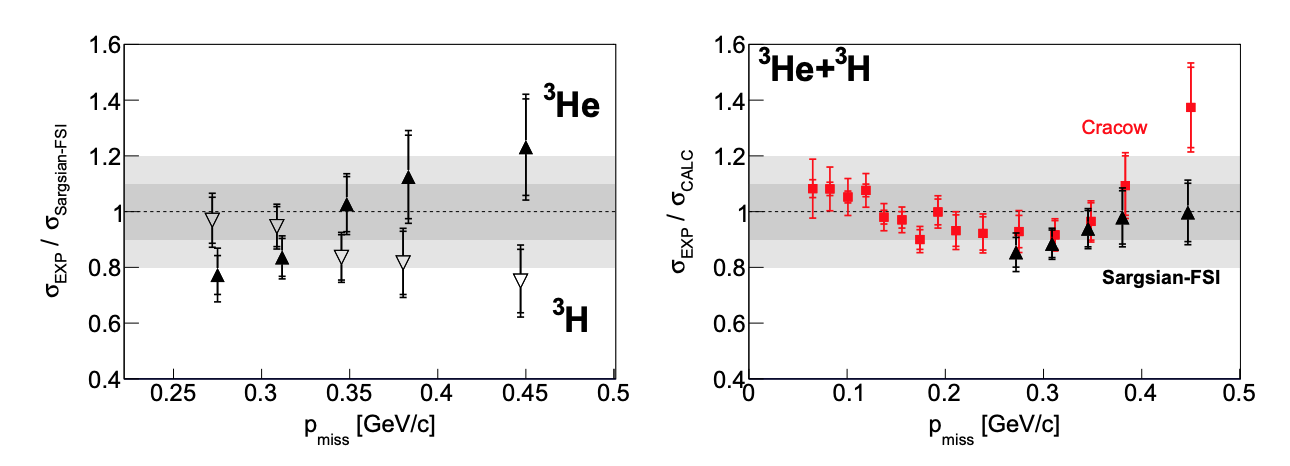}
\caption{Left: The ratio of the experimental cross
sections to the calculation of Sargsian that includes FSI of the leading nucleon for $^3$He (filled upright triangles) and $^3$H (open inverted triangles). Right: the ratio of the measured total $^3$He+$^3$H cross section to the Cracow PWIA calculation (red squares) and the Sargsian calculation that includes FSI (black triangles). The inner and outer bars show the statistical and statistical plus systematic uncertainties respectively. On both panels the shaded
regions show 10\% and 20\% agreement intervals.}
\label{fig:resultsIsoSum}
\end{centering}
\end{figure}

Fig~\ref{fig:resultsIsoSum} left shows the ratios of the measured cross sections  to the calculation of Sargsian that includes rescattering of the leading nucleon.  Including the effects of re-scattering of the outgoing nucleon improves the agreement between data and calculations at $p_{miss}>250$~MeV/c.  The diverging trend of this ratio for the two nuclei at higher \pmiss{} is possibly the result of single charge-exchange (SCX) re-scattering (e.g., $(e,e'n)$ neutron knockout followed by a $(n,p)$ charge exchange reaction) which could increase the \het\eep{} cross section and decrease the \trit\eep{} cross section.  

The isoscalar sum of $^3$He plus $^3$H (see Fig.~\ref{fig:resultsIsoSum})~\cite{Cruz-Torres:2020uke} should be largely insensitive to SCX.   Fig.~\ref{fig:resultsIsoSum} right shows the data to theory ratio for  the isoscalar sum of the $A=3$ nuclei.  Isoscalar data and theory agree to  within the uncertainty of the data. This validates current models of the ground state of $A=3$ nuclei up to very high initial nucleon momentum of 500~MeV/c. 

\begin{figure}[htb]
\begin{centering}
\includegraphics[width =0.55\textwidth]{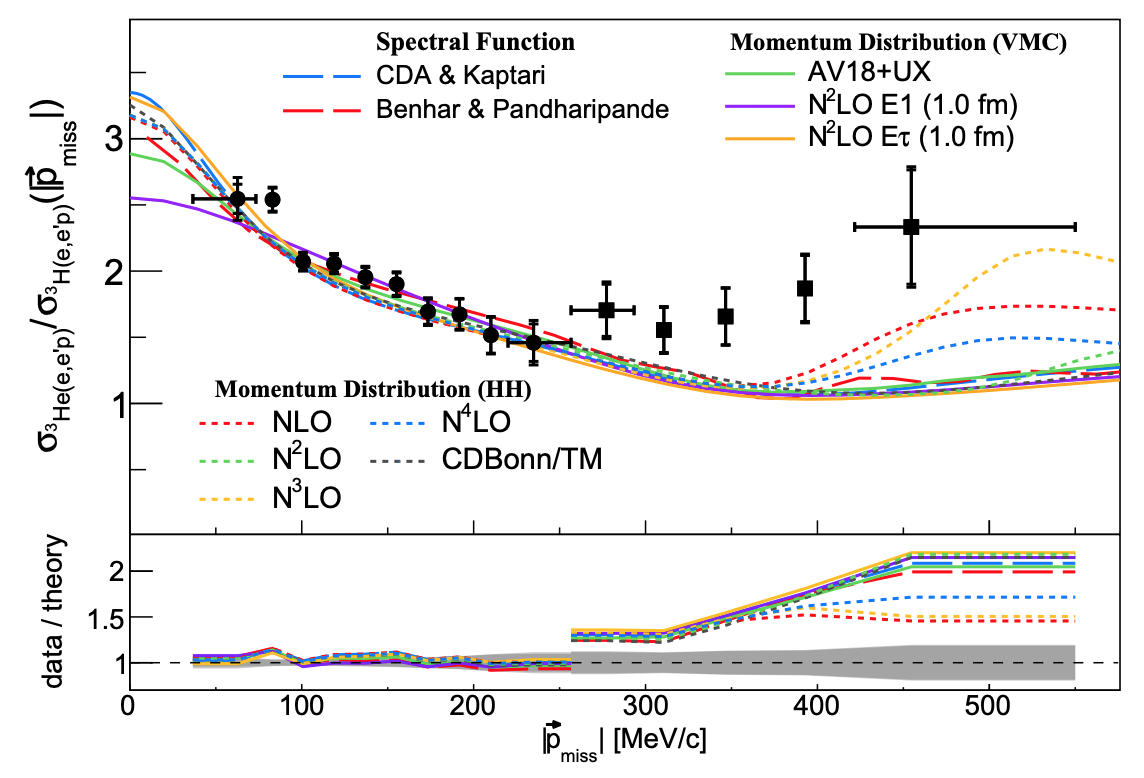}
\caption{The measured \het{} to \trit{} cross-section ratio,  $\sigma_{\het\eep} / \sigma_{\trit\eep}  (p_{miss})$,  plotted vs. \pmiss{} compared with different models of  the corresponding momentum distribution ratio~\cite{Cruz-Torres:2019bqw}. The filled circle and square markers correspond to the low and high \pmiss{} settings, respectively. Uncertainties shown include both statistical and  point-to-point systematical uncertainties. The overall normalization uncertainty of about $1.8\%$ is not shown. Horizontal bars indicate the bin sizes and are shown for only the first and last points in each kinematical setting as all other points are equally spaced. The bottom panel shows the double ratio of data to different calculated momentum distribution ratios, with the grey band showing the data uncertainty. The theoretical calculations are done using different local and non-local interactions, as well as different techniques for solving the three-body problem. }
\label{fig:ratio_halla}
\end{centering}
\end{figure}

Fig.~\ref{fig:ratio_halla} shows  the ratio of \het{} and \trit{} cross sections which should equal the ratio of the momentum distributions. In the simplest picture, the ratio of the number of protons in \hets to \trits should equal 2 at low momentum, due to simple nucleon counting, and decrease to 1 at high momentum, due to simple $np$ SRC pair counting. The ratio at low momentum is larger than 2, due to a shift of more low-momentum protons in \trits to high momentum caused by SRC pairing. This can be seen in the ratio of momentum distributions plotted in Fig.~\ref{fig:ratio_halla}.  The ratio of measured \het\eep{} to \trit\eep{} cross sections~\cite{Cruz-Torres:2019bqw} follows the calculated ratio of momentum distributions from $\pmiss=50$ to 250 MeV/c, but then is greater than that ratio by about 20--50\% at larger \pmiss. This is probably due to the SCX effects discussed above. Our proposed measurement will extend the measured $\pmiss$ spectra from the maximum of 0.5~GeV/c in the Hall A experiment to 1~GeV/c. We will also obtain high statistics over a range of angles that will enable detailed studies of the kinematic dependencies of the measurement and various FSIs.  

This proposal will improve on the Hall A experiment in several ways. The CLAS12 detector accesses significantly larger solid angle than the Hall A spectrometers. This substantial improvement in solid angle more than compensates for the decreased luminosity compared to the Hall A experiments and will dramatically improve both the statistical uncertainties and the range of \pmiss{} covered. It will also cover a wide range of kinematics (including a wide range in both $\theta_{rq}$, $x_B$) and will include studies of deuterium that will allow us to explore the effects of contributions from other reaction mechanisms in order minimize their effects.

\subsection{Asymmetric Nuclei}
\begin{figure}[htb]
\centering
\includegraphics[width = 0.6\textwidth]{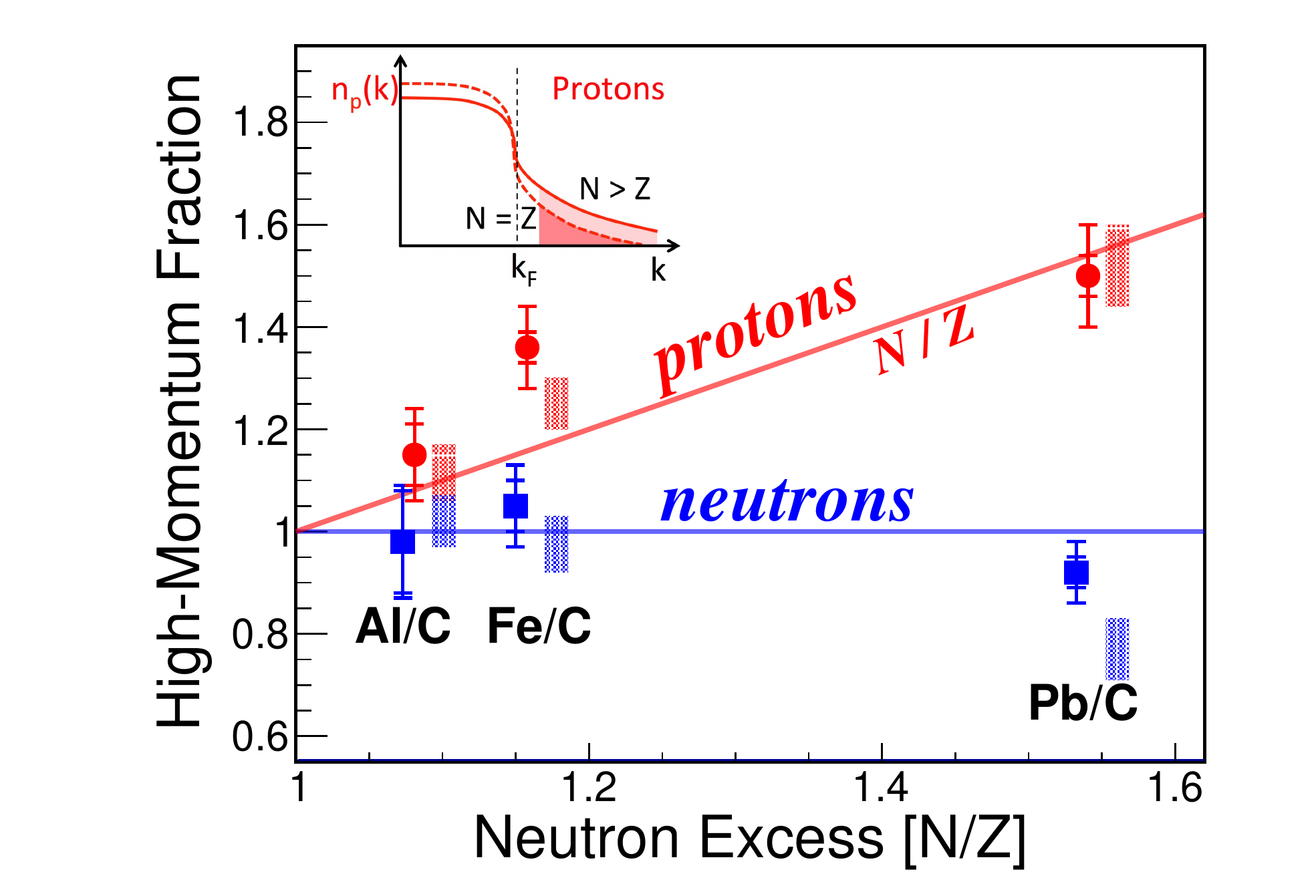}
\caption{Relative high-momentum fractions for neutrons and protons. Red (blue) circles are double ratio of the number of \eep{} high-momentum events to low-momentum proton (neutron) events for nucleus $A$ relative to $^{12}$C.}
\label{Duer-Pr-Ne}
\end{figure}

We previously re-analyzed CLAS data to extract the double ratio of the high-momentum fraction from different nuclei to $^{12}$C for both the proton and the neutron to show that the proton is more correlated in neutron-rich nuclei~\cite{Duer:2018sby}. This suggests that the minority nucleons (i.e., the protons) move faster than the majority nucleons (i.e., the neutrons) in neutron-rich nuclei, see Fig. \ref{Duer-Pr-Ne}. This result is opposite the simple expectation from a simple Fermi gas or mean-field nucleus. 

This result was further tested by measuring the ratio of inclusive \ee{} cross sections of $^{48}$Ca to $^{40}$Ca in Hall A~\cite{Nguyen:2020mgo}. This measurement tests the effect of adding eight neutrons on the high momentum nucleons in Ca. The measured $^{48}$Ca/$^{40}$Ca cross section ratio of about 1.17 shows that increasing the number of neutrons increases the fraction of high-momentum nucleons and thus increases the number of correlated pairs. This is consistent with the observation of Ref.~\cite{Duer:2018sby}, but because it is an inclusive measurement, we can not separate the proton and neutron contributions. The upcoming CaFe experiment~\cite{proposal_cafe} in Hall C will measure \eep{} on $^{40}$Ca and $^{48}$Ca to answer the questions: 1) Does $^{48}$Ca have more protons in SRCs compared to $^{40}$Ca? 2) What is the high-momentum fraction of protons in $^{48}$Ca?

All of these studies focus on medium and heavy asymmetric nuclei. It will be interesting to extend these studies to light asymmetric nuclei, and the most asymmetric stable light nuclei are \trits ($N/Z =2$) and \hets ($N/Z=0.5$). Particularly, these nuclei will test theory and our understanding of the observation we have made in the heavier nuclei. In this proposal, we will extract the high-momentum fraction for both protons and neutrons for \trits and \hets and compare them to the results obtained from heavier asymmetric nuclei. 

\subsection{Short range correlations}

Studies of the short-ranged structure of nucleons in the nucleus have produced many fascinating results (see Ref.~\cite{Hen2016kwk} and references therein). Per-nucleon ratios of inclusive electron scattering cross sections indicate that all nuclei have similar momentum distributions for $p_i\ge 275\pm25$ MeV/c and that this is due to the effects of $NN$ SRC pairs.  Measurements on $^{12}$C and $^4$He using both the $(e,e'pN)$ and $(p,2pn)$ reactions show that knock-out of a high-initial-momentum ($p_i\ge 300$ MeV/c) proton from the nucleus almost always results in the emission of its correlated partner nucleon and that nucleon is almost always a neutron (for $300\le p_p \le 600$ MeV/c). This indicates that almost all of these high-momentum nucleons belong to an SRC pair and that these pairs are predominantly $pn$ pairs. These results were confirmed for heavier nuclei using data from CLAS~\cite{Hen2014nza,Duer:2018sxh,Duer:2018sby}. These SRC pairs have a center-of-mass momentum distribution that is consistent with adding the momenta of two mean-field nucleons~\cite{Cohen:2018gzh}, and the probability of finding an SRC pair in a nucleus is proportional to the probability that two nucleons are in a node-less relative $S$-state~\cite{Colle2015ena}.  This indicates that SRC pairs are momentary fluctuations of two short-distance nucleons into a short-lived high-relative momentum state. 

\begin{figure*}[htb]
\includegraphics[width = 6.5cm]{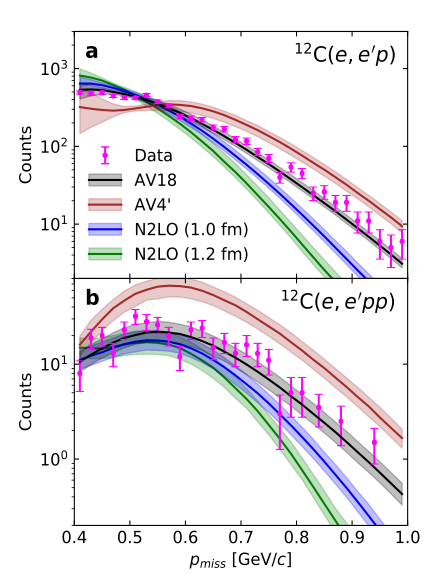}
\includegraphics[width = 7.5cm]{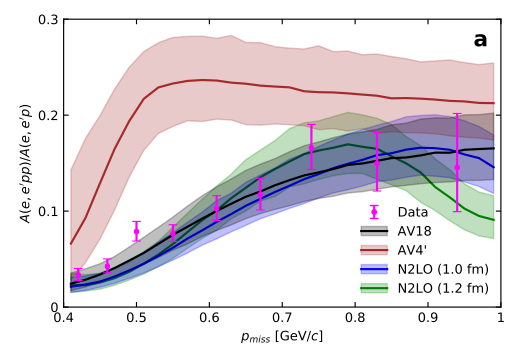}
\caption{(left) Measured 12C(e,e'p) (a) and 12C(e,e'pp) (b) event yields shown as a function of the (e,e’p) missing momentum  compared with theoretical calculations based on the GCF framework, using different models of the NN interaction; (right) Measured 12C\eepp/\eep{} event yields ratios shown as a function of the \eep{} missing momentum compared with theoretical calculations based on the GCF framework using different models of the $NN$ interaction~\cite{Schmidt:2020kcl}. }
\label{fig:Nature_GCF}
\end{figure*}

\begin{figure*}[htb]
\includegraphics[width = 6.5cm]{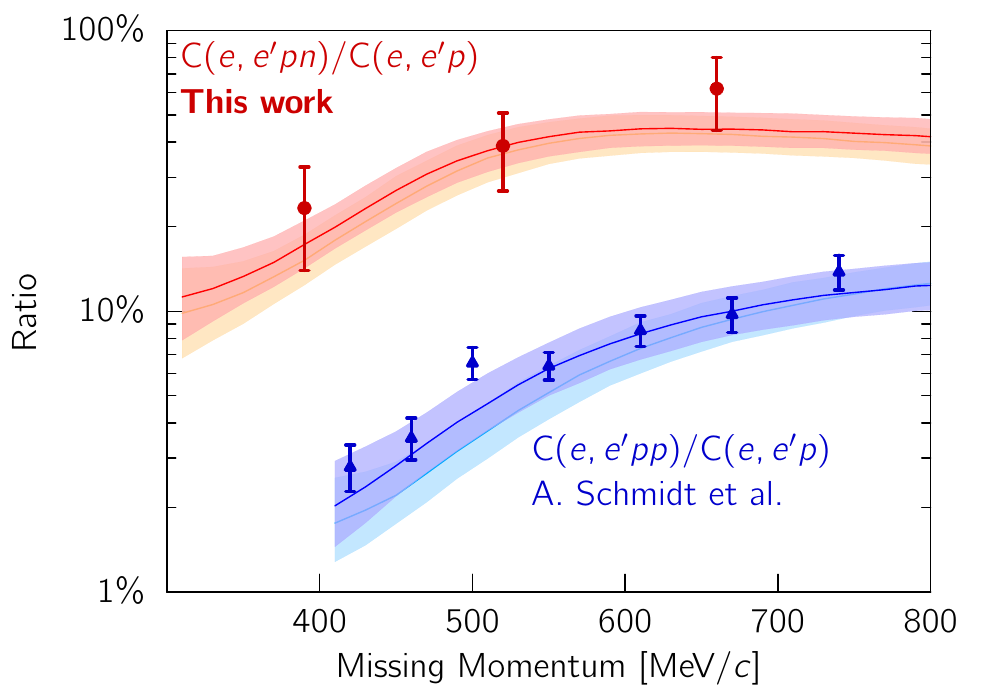}
\includegraphics[width = 6.5cm]{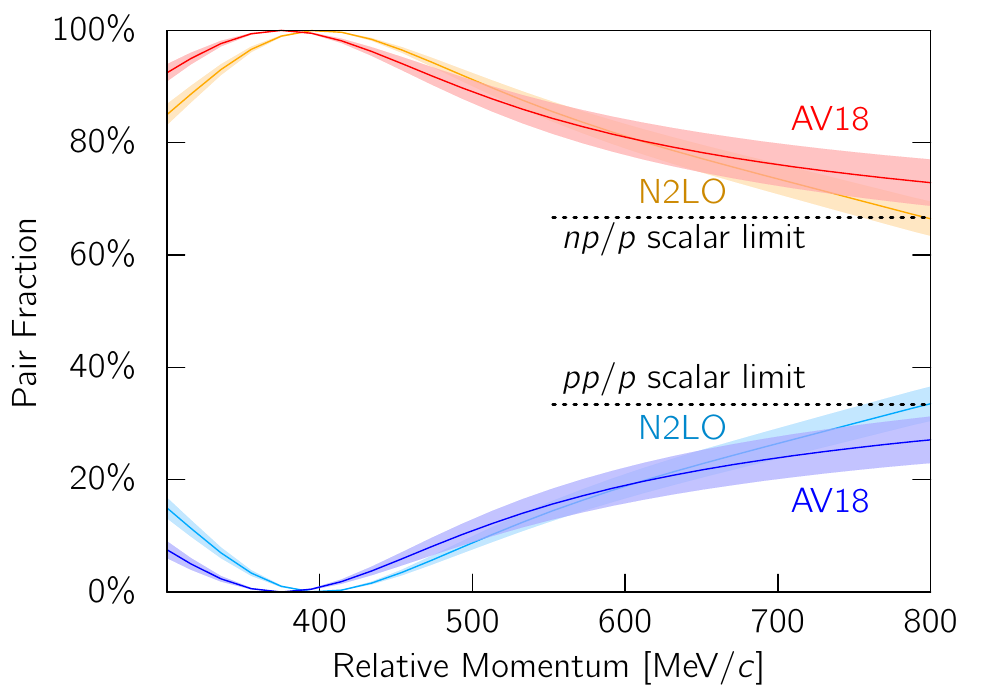}
\caption{Left: The measured fractions of triple coincidence events (C$(e,e'pN)/$C$(e,e'p)$), compared with GCF predictions accounting for the variety of effects that influence the measurement (e.g. CLAS detector acceptance, efficiency, and resolution, FSIs including SCX, and the event-selection procedure) \cite{Korover:2020lqf}. The C$(e,e'pp)/$C$(e,e'p)$ data (blue triangles) are taken from Ref.~\cite{Schmidt:2020kcl}, while the C$(e,e'pn)/$C$(e,e'p)$ data (red circles) are from this work. Right: The GCF prediction for the ground-state fractions of $pn$ and $pp$ pairs as a function of pair relative momentum, calculated using the AV18 and N2LO $NN$ interactions. The dashed line marks the scalar limit. The width of the GCF calculation bands shows their 68\% confidence interval due to uncertainties on the model parameters.}
\label{fig:korover_GCF}
\end{figure*}

More recently, measurements of \eep, \eepp, and $(e,e'pn)$ up to much higher nucleon momentum~\cite{Schmidt:2020kcl,Korover:2020lqf} show a transition from $300\le p_i\le 600$ MeV/c, where there are far more $np$ than $pp$ pairs, to $600\le p_i\le 1000$ MeV/c, where the relative number of $np$ and $pp$ pairs is determined by simple counting (see Figs.~\ref{fig:Nature_GCF} and \ref{fig:korover_GCF}).  This shows the transition from a spin-dependent (tensor) $NN$ interaction to a spin-independent (scalar) interaction at high momentum.

We can describe these SRC pairs using the newly developed generalized contact formalism (GCF). The GCF exploits the scale separation between the strong interaction between the nucleons in an SRC pair and the pair's weaker interaction with its surroundings~\cite{Weiss:2015mba,Weiss:2016obx,Cohen:2018gzh}. Using this scale separation, the two-nucleon density in either coordinate or momentum space (\textit{i.e.}, the probability of finding two nucleons with relative and c.m. momenta $q$ and $Q$ respectively, or with separation $r$ and distance $R$ from the nucleus c.m.~\cite{Wiringa:2014}) can be expressed at small separation or high relative momentum as~\cite{Weiss:2016obx}:
\begin{align}
	\rho_{\alpha, NN}^{A}(R,r) & = C_{\alpha, NN}^{A}(R) \times |\varphi_{NN}^{\alpha}(r)|^2 , \nonumber \\
	   n_{\alpha, NN}^{A}(Q,q) & = \tilde{C}_{\alpha, NN}^{A}(Q) \times |\tilde\varphi_{NN}^{\alpha}(q)|^2 , 
\label{Eq1}
\end{align}
where $A$ denotes the nucleus, $NN$ the nucleon pair ($pn$, $pp$, $nn$), 
and $\alpha$ stands for the quantum state (spin 0 or 1). $\varphi_{NN}^{\alpha}$ are universal two-body wave 
functions, given by the zero-energy solution of the two-body Schr\"odinger equation, and $\tilde\varphi_{NN}^{\alpha}$ are their Fourier transforms. 
$\varphi_{NN}^{\alpha}$ are universal in the weak sense, \textit{i.e.}, they are nucleus independent but not model independent.
Nucleus-dependent  ``nuclear contact coefficients'' are given by
\begin{align}
	C_{\alpha, NN}^{A} & \equiv \int d{\bf R} \; C_{\alpha, NN}^{A}(R), \nonumber \\
	\tilde{C}_{\alpha, NN}^{A} & \equiv \frac{1}{(2\pi)^3} \int d{\bf Q} \; \tilde{C}_{\alpha, NN}^{A}(Q),
\label{Eq2}
\end{align}
and define the number of $NN$-SRC pairs in nucleus $A$.

The GCF describes the measured momentum distributions of nuclei for \eep, \eepp{} and $(e,e'pn)$ reactions remarkably well (see Figs.~\ref{fig:Nature_GCF} and \ref{fig:korover_GCF}) and allows us to test the predictions of different $NN$ interactions.  By measuring \eep{} and \eepp{} up to $\pmiss=1000$ MeV/c we can test $NN$ interactions in nuclei at remarkably short distances. 

The $NN$ interaction is a crucial input for calculations of nuclear structure and reactions as well as for other studies such as neutrino-less double beta decay and neutron stars. The $NN$ force is not a fundamental force; it is due to the "leakage" of the strong interaction that binds quarks together to form the nucleon. Therefore $NN$ interactions are described by effective theories. Current models have limited predictive power and are  loosely constrained at short distance. Measuring nucleon momentum distributions in nuclei to high momenta allows us to constrain the $NN$ interaction at previously unreachable short distances.


In Fig.~\ref{fig:Nature_GCF}, the measured ratio for $^{12}$C\eepp/\eep{} is compared with GCF calculations using the phenomenological AV18 interaction, $\chi$EFT next-to-next-to-leading order (N2LO) interactions, and the scalar-only AV4' (lacking the tensor force) interaction. The AV18 potential agrees well with the data. The N2LO potentials include explicit cut offs at distances of 1 and 1.2 fm, corresponding to momentum cut offs at approximately 400-500~MeV/c and do not describe the data well above this cut off, as expected. The AV4' interaction is scalar-only (lacking the tensor force) and agrees with data in the scalar-dominated high-momentum region but fails in the tensor-dominated low-momentum region~\cite{Schmidt:2020kcl}.

By extending these measurements to few-body nuclei we gain several advantages. Few-body nuclei already exhibit the same range of complex nuclear phenomena, including $NN$ SRC pairs, as heavy nuclei, but they are far easier to calculate. There are many calculations of the ground state of $A=3$ nuclei \cite{Ford_2014}. We can also compare these predictions to the more approximate GCF predictions, which can more easily incorporate many different $NN$ interactions. The effects of FSI are much smaller in \eep{} and \eepN{} on few body nuclei, because there are many fewer nucleons to re-scatter from. There are calculations of nucleon re-scattering in \het\eep{} as discussed above. In addition, $A=3$ nuclei have the added advantage of having both the largest and the smallest neutron to proton ratios of any "stable" nucleus with $A>1$. This will allow us to test $np$ and $pp$ pairing hypotheses in the most extreme systems available.

We will measure the fraction of \eep{} high-\pmiss{} events with an associated second nucleon in order to study  SRC pairing in $A=3$ nuclei.  If the struck proton belonged to an SRC pair, then its partner nucleon should be ejected at high momentum and the 3rd, spectator nucleon, will have lower momentum $\vec p_3=\vec p_{cm}$ where $\vec p_{cm}$ is the center of mass momentum of the correlated pair. By looking at how the fraction of \eep{} events with a correlated partner proton grows (and how the fraction with a correlated neutron partner decreases) as \pmiss{} increases from 300 to 1000 MeV/c, we can study the transition from the tensor-dominated to the scalar-dominated part of the $NN$ interaction. While the spin-1 $pn$ pairs are dominant, this experiment will also let us study the 20-times less common spin-0 $pp$ pairs. We will also explore other open questions, including  the cm momentum distribution of $pp$ and $pn$ pairs, the relative momentum distribution of $pp$ and $pn$ pairs, and the relationship between the relative and cm momentum. All of these quantities are exactly calculable in $A=3$ nuclei (for a given $NN$ potential), unlike in heavier nuclei.

\subsection{Neutron magnetic form factor}

Electromagnetic form factors are fundamental, measurable quantities of nucleons describing their charge and current distributions as functions of $Q^2$. The form factors are essential for constraining nucleon models, understanding spontaneous symmetry breaking in QCD, and constitute the zeroth moment of generalized parton distributions. Due to their importance in constraining our understanding of the nucleonic picture, the nucleon electromagnetic form factors are a key experimental objective of the JLab 12 GeV program with a goal being to measure the form factors over a large range of $Q^2$ with high-precision. 

The lack of a free neutron target poses special challenges for the extraction of the form factors of the neutron. Neutron form factors must be extracted from quasi-elastic or elastic scattering measurements on deuterium or light nuclei, taking into account the effect of the nuclear wave function. Cross section measurements are essentially only sensitive to the neutron's magnetic form factor, $G_M^n$, since the neutron's electric form factor, $G_E^n$ is much smaller by comparison, and can only be accessed through polarization asymmetries, sensitive to the $G_E^n/G_M^n$  ratio. 

Most previous determinations of $G_M^n$ have been made from quasi-elastic scattering cross sections on deuterium. Systematic improvements can be made by tagging the struck nucleon, i.e., through the $d(e,e'n)p$ reaction, and further by simultaneously comparing to the $d(e,e'p)n$ reaction. The most precise determination of $G_M^n$ was made over the $Q^2$ range from 1 to 5~[GeV/c]$^2$ using this technique at CLAS~\cite{Lachniet_2009}. In the 12-GeV era, the CLAS-12 Run Group B recently collected data and intends to extract $G_M^n$ to much larger $Q^2$, and improve on the precision of earlier SLAC measurements~\cite{PhysRevD.46.24}. The Super-Big Bite (SBS) program in Hall A intends to measure $G_M^n$ out to $Q^2=13.5$~[GeV/c]$^2$.

\begin{figure}[htb]
\begin{center}
\begin{minipage}{0.7\textwidth}
\includegraphics[width=\textwidth]{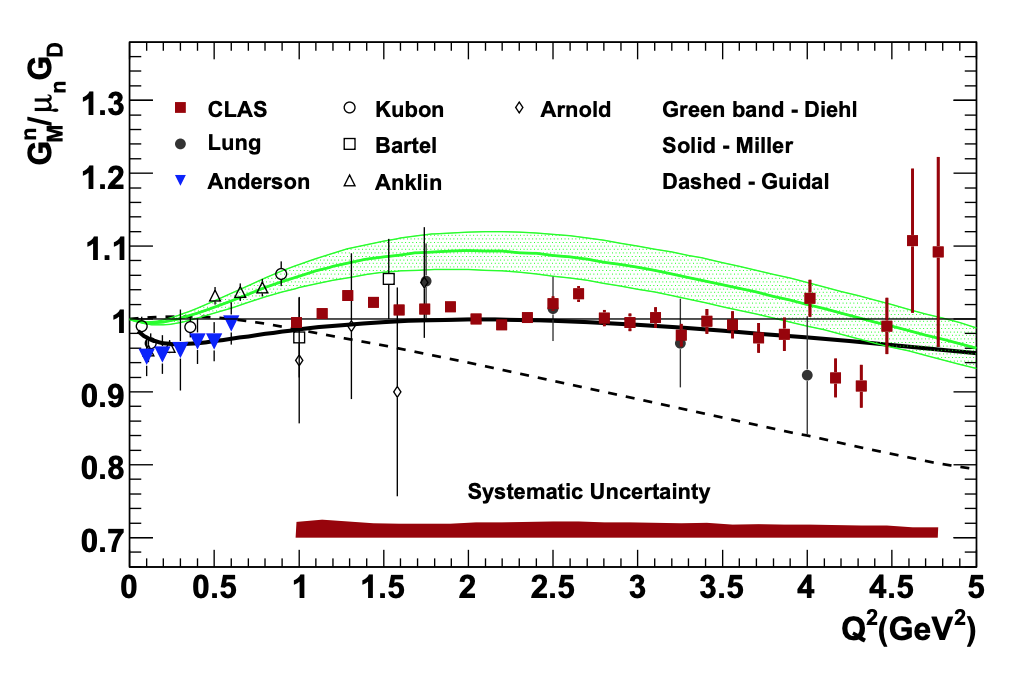}
\end{minipage}
\end{center}
\caption{Select data for G$_M^n$ at low $Q^2$ where the red points from CLAS cover the largest $Q^2$ range with good precision~\cite{Lachniet_2009}. There exists significant contention between the data and model predictions at $Q^2\le 0.5$[GeV/c]$^2$.}
\label{fig:GMn_range}
\end{figure}

Despite the push to higher $Q^2$, there is still a troubling amount of uncertainty on $G_M^n$ at low $Q^2$. A selection of previously measured $G_M^n$ values for $Q^2<5$~[GeV/c]$^2$ is shown in Fig.~\ref{fig:GMn_range}, relative to the standard dipole form factor, $G_D=(1+Q^2/\Lambda^2)^{-2}$, where $\Lambda^2=0.71$~GeV$^2/c^2$. Below 1~[GeV/c]$^2$, there is a discrepancy between older measurements~\cite{Markowitz:1993hx,Bruins:1995ns}, and more recent measurements~\cite{Anklin:1998ae,Kubon:2001rj} which found $G_M^n$ to be slightly larger. A subsequent measurement in Hall A of the transverse beam-target asymmetry on polarized $^3$He found a smaller $G_M^n$~\cite{Anderson:2006jp}. This low-$Q^2$ discrepancy persists in theory, with Cloudy Bag Model calculations~\cite{Miller:2002ig} favoring the larger $G_M^n$, while GPD-based calculations favor smaller values~\cite{Guidal:2004nd,Diehl:2004cx}. New data are needed to help pin down $G_M^n$ below 1~[GeV/c]$^2$, preferably with different systematic uncertainties.

The use of a tritium target presents a clear opportunity to make a defining low-$Q^2$ measurement of $G_M^n$. By measuring the inclusive quasi-elastic cross section for scattering from tritium relative to helium-3, one can extract $G_M^n$ relative to the proton's magnetic form factor $G_M^p$, which is much better known. In the limit where nucleons are stationary, the ratio of $G_M^n/G_M^p$ can be written
\begin{equation}
    \left(\frac{G_M^n}{G_M^p}\right)^2 = \frac{\left(2\frac{\sigma_{^3H}}{\sigma_{^3He}}-1\right)\left[1+\frac{\epsilon}{\tau}\left(\frac{G_E^p}{G_M^p}\right)^2 \right]}{2+\frac{\sigma_{^3H}}{\sigma_{^3He}}}, 
\end{equation}
where $\tau\equiv Q^2/4m_N^2$, and $\epsilon$ is the virtual photon polarization parameter. The Fermi-motion of the nucleons in the $A=3$ wave function complicate this simple relationship but do not diminish the sensitivity of the cross section ratio to $G_M^n$.  

Extracting $G_M^n$  by measuring \trit\ee{} and \het\ee{} carries the huge advantage that no neutron detection is necessary. The inclusive measurements also count much faster than  $(e,e'N)$ measurements. Furthermore, the sensitivity to the nuclear wave function is reduced by taking a cross section ratio. In fact, an extraction of $G_M^n$ through this technique (at higher $Q^2$) was one of the main components of the 2018 Hall A experiment E12-11-112 is using this technique to measure $G_M^n$ for $0.35\le Q^2\le 2.75$ (see Fig.~\ref{fig:GMn_reach}). The CLAS measurement has the advantage of  simultaneously measuring a higher density of points and a larger range of $Q^2$, where high resolution is not critical. 

This proposed experiment will measure the inclusive quasi-elastic cross section ratio of $^3$H$(e,e')$ / $^3$He$(e,e')$ to make a high precision determination of $G_M^n$ focusing on the region of $Q^2<1$~[GeV/c]$^2$. This will shed valuable light on the discrepancy in previous low-$Q^2$ extractions of $G_M^n$ with very different systematic uncertainties. The measurement will also complement the CLAS12 Run Group B and Hall A SBS measurements, allowing better constraint of the models of the nucleon form factors over the complete momentum transfer range. 

\section{Physics Goals} \label{physGoals}

We will  significantly impact our interpretation of models and constrain the theory of fundamental few-body nuclear physics by measuring cross sections with high statistics over a large kinematic range on both $^3$He and $^3$H targets. The isospin asymmetry of these two targets will also enable us to further extract information about the momentum distributions of minority and majority nucleons and the effects of SRC pairs. Through this measurement, we will specifically:  
\begin{itemize}
    \item benchmark few-body nuclear models,
    \item constrain the $NN$ interaction and the nuclear wave function at high momentum,
    \item study scale separation in SRC pairs and pair formation mechanisms,
    \item determine the isospin  of SRC pairs at different momenta,
    \item measure $G_M^n$ at low and moderate $Q^2$
\end{itemize}

All of these goals are crucial for our understanding and interpretation of the dynamics in nuclei and will refine theory predictions for heavier nuclear systems. The inclusion of deuterium data will complement the measurements on \het{} and \trit{} and are critical to the evaluation of non-QE contributions in the measured cross sections and observables.  (Note that both the $NN$ interaction and wave function are model dependent quantities.  Unitary transformations can shift strength from the operator to the wave function and vice versa.)

\subsection{$NN$ interaction}
\begin{figure}[htb]
\centering
\includegraphics[width=0.7\textwidth]{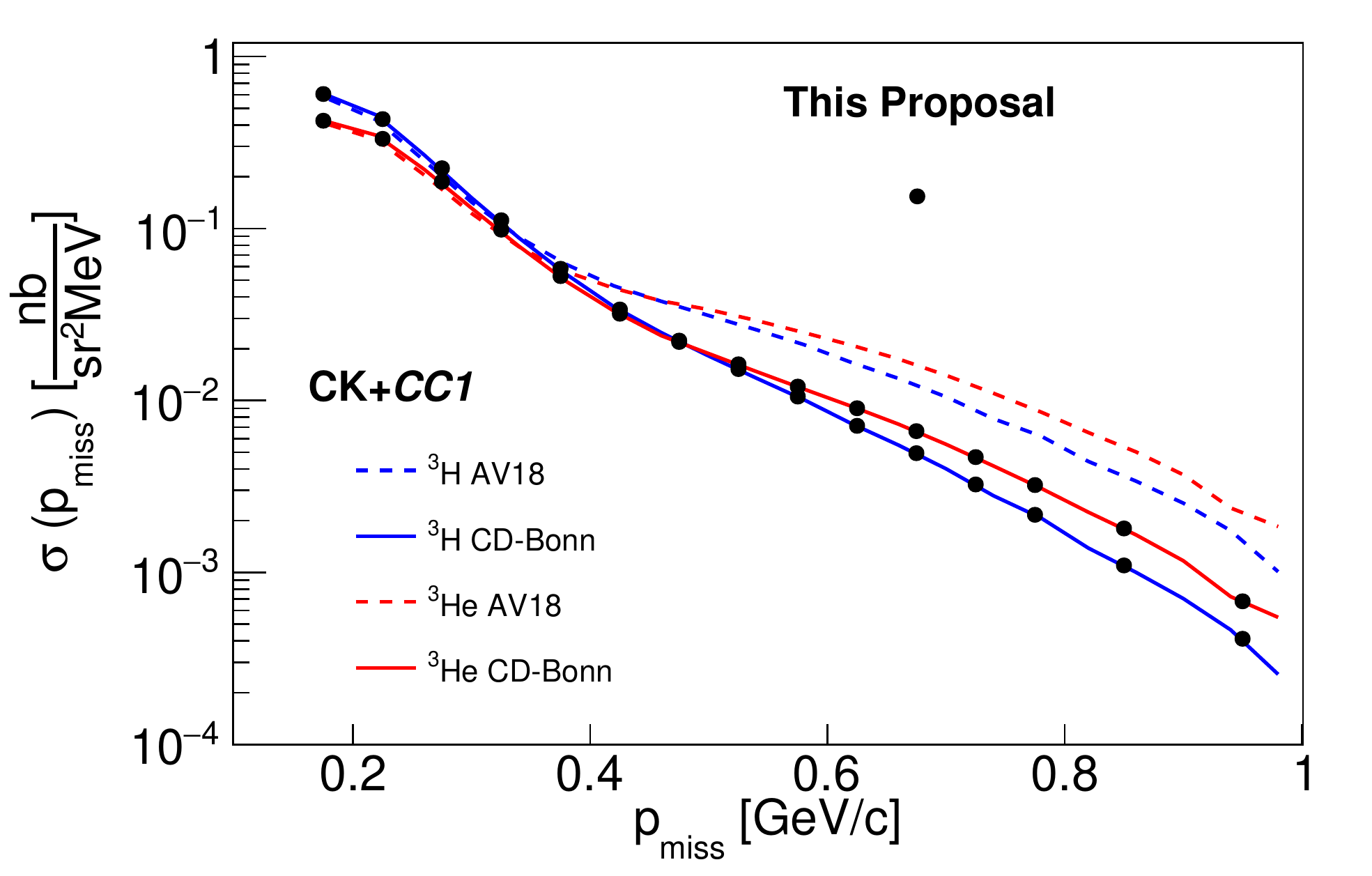}
\caption{The factorized absolute cross section calculations using the  Ciofi degli Atti and Kaptari spectral function together with the DeForest CC1 off-shell cross section are shown  for both \het\eep{} and \trit\eep{} for the AV18 and CD-Bonn $NN$ interactions, separately. Our projected data points are shown by the black circles including our estimated statistical uncertainty and a 5\% point-to-point systematic uncertainty (which are smaller than the data points).} 
\label{fig:CK-XS_pred}
\end{figure}
\begin{figure}[htb]
\includegraphics[width=0.45\textwidth]{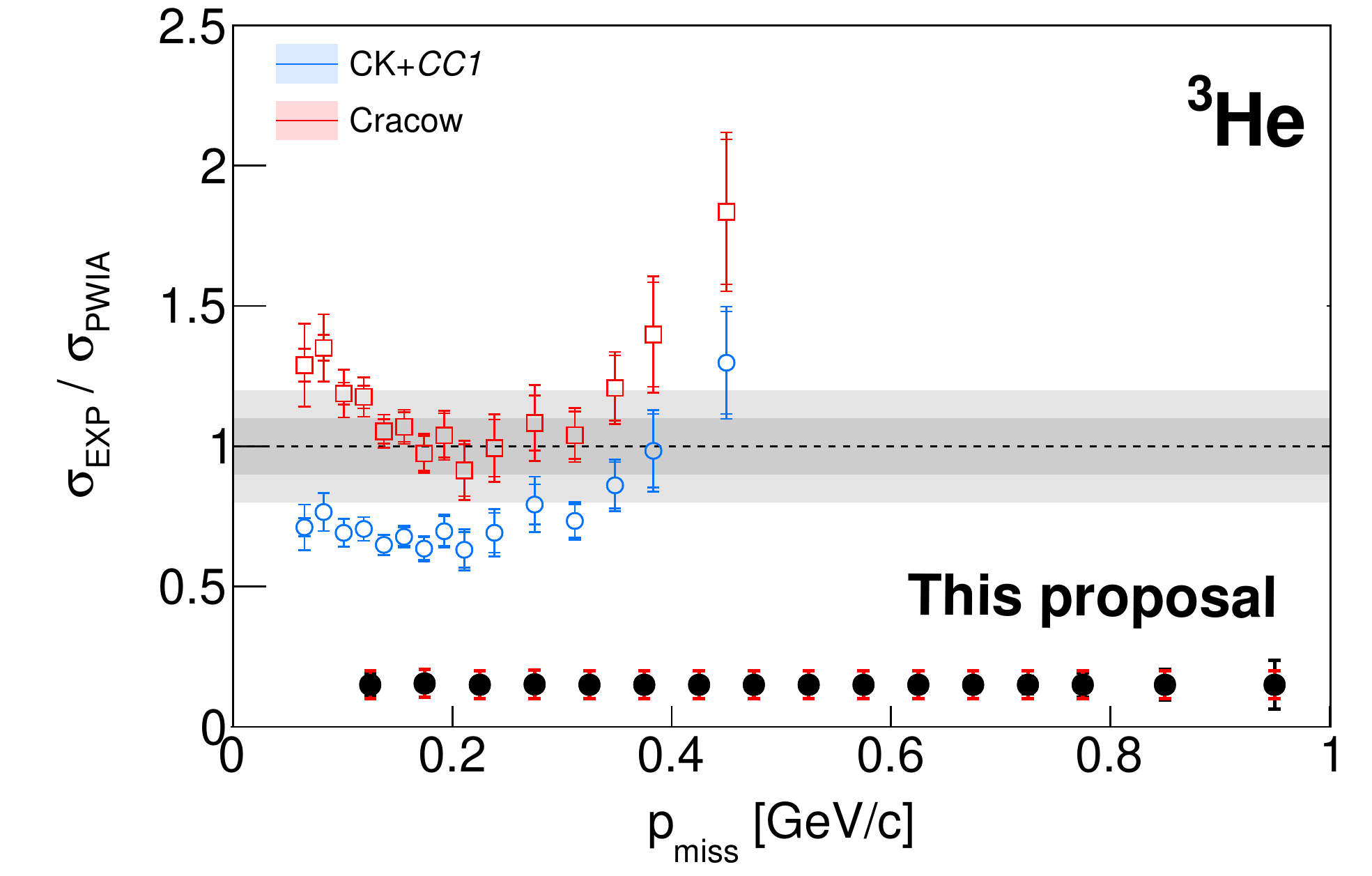}
\includegraphics[width=0.45\textwidth]{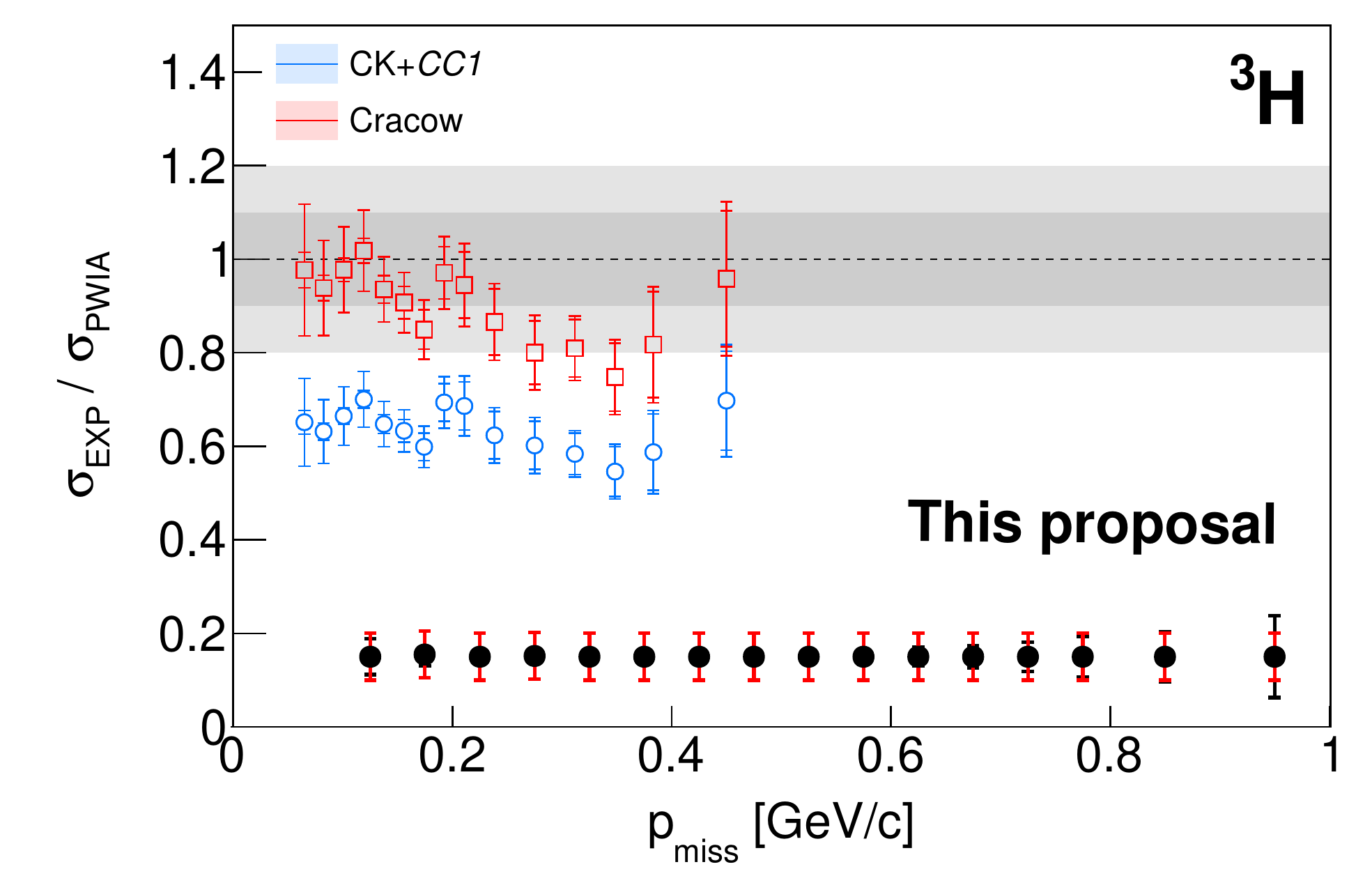}
\caption{Left: The \het\eep{} experimental cross sections from Ref.~\cite{Cruz-Torres:2020uke} are shown with our proposed measurement (black triangles with statistical (black) and a 5\% point-to-point systematic (red)  uncertainty - the uncertainties are not combined). Right: The \trit{} experimental cross sections from Ref.~\cite{Cruz-Torres:2020uke} are shown with our proposed measurement.}
\label{fig:A3-proj}
\end{figure}
\begin{figure}[htb]
\centering
\includegraphics[width=0.7\textwidth]{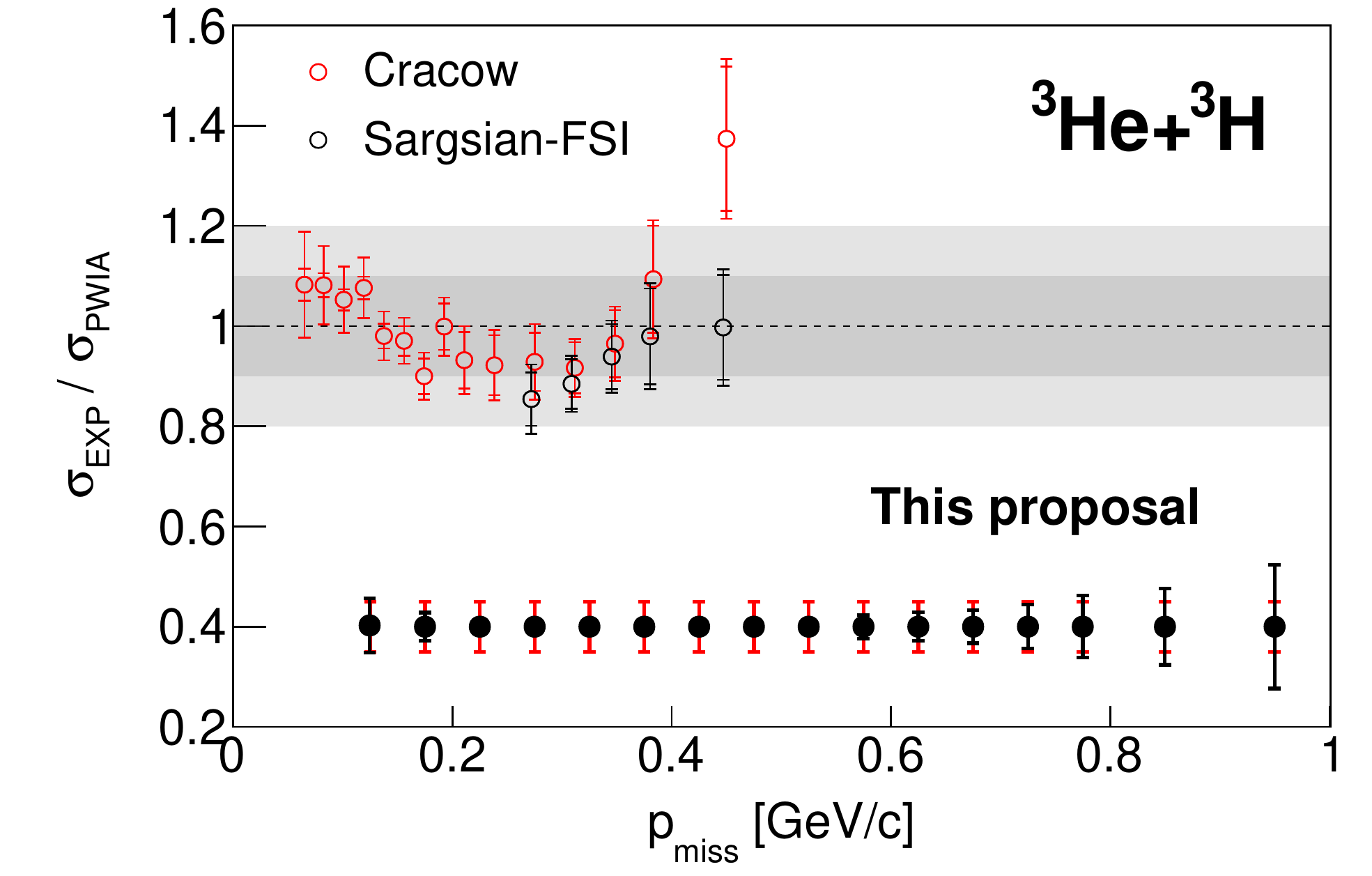}
\caption{The ratio of the measured total $^3$He+ $^3$H cross sections relative to the Cracow PWIA and the Sargsian calculation that include FSI described in Ref.~\cite{Cruz-Torres:2020uke}. The projection of our measurement is shown by the black solid points where black error bar and red error bars are statistical and a 5\% point-to-point systematic uncertainty, respectively.}
\label{fig:sum-proj}
\end{figure}

We will measure QE absolute cross sections for \eep{} on both $^3$He and $^3$H  to constrain $NN$ interaction models. We will measure over a wide range of $x_B$ and $Q^2$ with \pmiss{} up to $\approx 1$~GeV/c. The \eep{} cross sections will be compared to nuclear theory predictions using a wide variety of techniques and $NN$ interactions in order to constrain the $NN$ interaction at short distances. 

Fig.~\ref{fig:CK-XS_pred} shows a factorized calculation of the absolute \het\eep{} and \trit\eep{} cross sections using the 3He spectral function of
C. Ciofi degli Atti and L. P. Kaptari including the continuum interaction of the two spectator nucleons \cite{degli_Atti_2005} and the $\sigma_{cc1}$ electron off-shell nucleon cross section \cite{DeForest:1983ahx} using both the AV18 \cite{Wiringa_1995} and CD-Bonn \cite{Machleidt_2001} $NN$ interactions. Due to the lack of \trit{} proton spectral functions, we assume isospin symmetry and use the \het{} neutron spectral function. The expected uncertainties are smaller than the points.

Our cross section measurements will significantly extend the  Hall A tritium measurements. The cross sections measured in Hall A along with our projected measurement in this proposal are shown in Fig.~\ref{fig:A3-proj} and are compared to different PWIA calculations. 

The proposed measurements we describe are shown as the black triangles in Fig.~\ref{fig:A3-proj} along with the anticipated statistical and 5\% point-to-point systematic uncertainties. The estimated statistical uncertainty is based on the proposed running in Table~\ref{requestedbeam}. We will make significant contributions for all \pmiss{} up to 1~GeV/c.

The isoscalar sum of the \het{} and \trit{} cross sections compared to PWIA calculations are shown in Fig.~\ref{fig:sum-proj}. This sum reduces the contributions from SCX and improves our sensitivity in evaluating the $NN$ ground state. Furthermore, our measurement will be the first to evaluate calculations of these nuclei up to \pmiss{} of 1~GeV/c. 

\subsection{Formation mechanisms and isospin dependence of SRC pairs}
\begin{figure}[htb]
\centering
\begin{minipage}{0.45\textwidth}
\includegraphics[width=\textwidth]{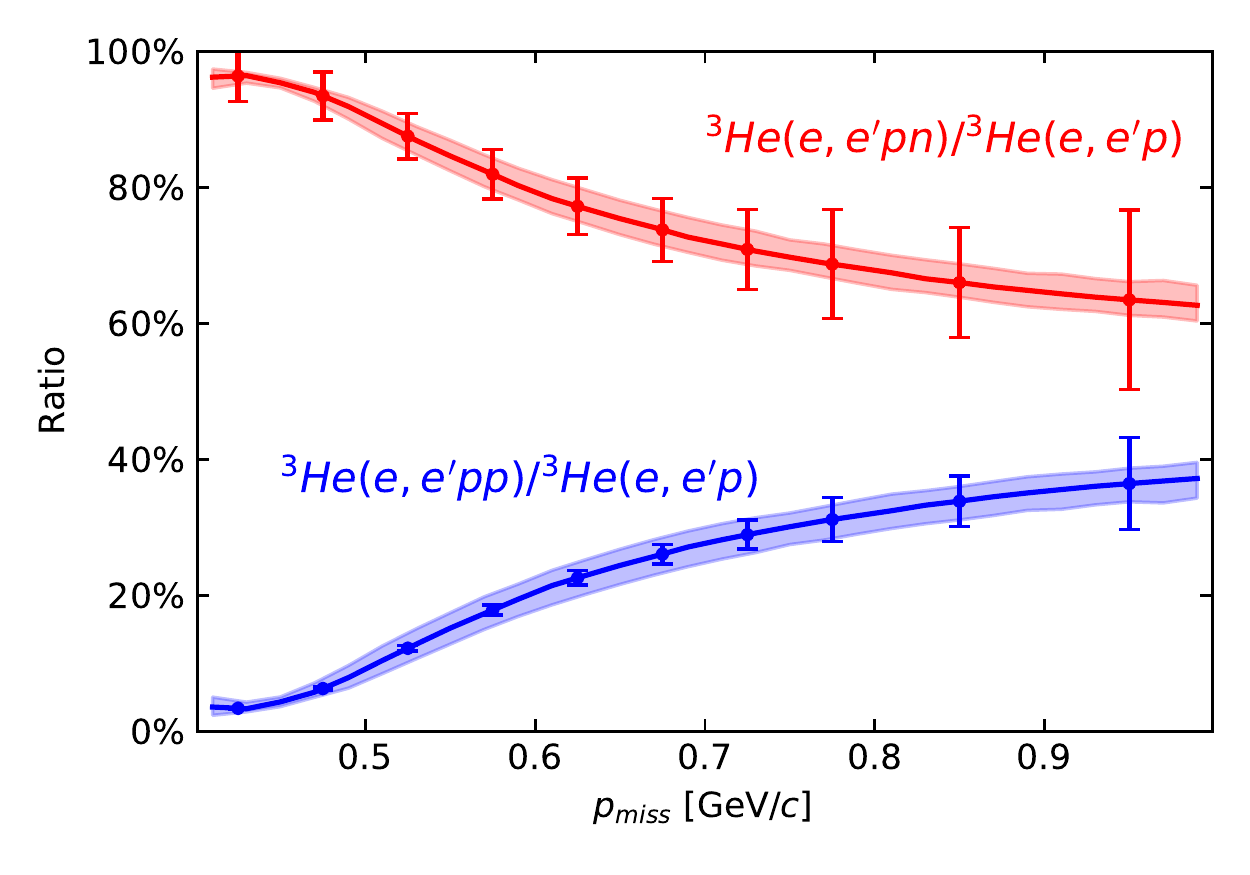}
\end{minipage}
\caption{The GCF prediction for \het{} ratios of \eepp/\eep{} and $(e,e'pn)$/\eep{} using the AV18 interaction, $\sigma_{c.m.}=100\pm 20$~MeV, and the contacts from Ref.~\cite{cruztorres2019scale}. The points indicate the projected measurements of this proposal with anticipated statistical and 5\% systematic uncertainty in the ratio. The \trit$(e,e'pn)$/\trit\eep{} ratio (not shown) is expected to be approximately 1.}
\label{fig:gcf_proj}
\end{figure}

We will measure the $(e,e'pN)$ quasi-elastic reaction cross sections and the $(e,e'pN)/\eep$ ratios to understand SRC $NN$ pairs in the simplest non-trivial system. 
For struck protons belonging to an SRC pair, the partner nucleon should be ejected at high momentum and the third, spectator nucleon, should have lower momentum $\vec p_3=\vec p_{cm}$ where $\vec p_{cm}$ is the center of mass momentum of the correlated pair. We will measure  how the fraction of \eepp/\eep events (the fraction of $pp$ SRC pairs) changes with \pmiss. This fraction should increase with increasing \pmiss{}  and show us the transition from the tensor to scalar-dominated regimes of the $NN$ interaction. Similarly, the \eepn/\eep fraction should decrease with \pmiss.

As described in the Section~\ref{overview}, we can exploit the scale separation utilized by the GCF which describes the measured momentum distributions of nuclei for \eep, \eepp{} and $(e,e'pn)$ reactions and makes predictions for different $NN$ interactions. In this proposal, we will extract the contact terms for \het{} and \trit. Using the GCF, we can predict the \eepp/\eep{} and $(e,e'pn)$\eep{} cross section ratios using different $NN$ interactions as shown in Fig.~\ref{fig:gcf_proj}.

The advantage of measuring $A=3$ nuclei versus heavier nuclei as in many previous SRC observations is that the characteristics of these nuclei are exactly calculable. As shown in Fig.~\ref{fig:gcf_proj}, we will extend the missing momentum range probing the $NN$ interactions in nuclei at extremely short distances. The spin-1 $pn$ are dominant at high \pmiss{}, but this experiment will uniquely enable us to explore the 20-times less common spin-0 $pp$ pairs. We can measure the center-of-mass momentum distributions of the $pp$ and $pn$ pairs, the relative momentum of the $pp$ and $pn$ pairs, and we can quantify the relationship between the relative and center-of-mass momentum. Importantly, all of these quantities are precisely calculable in  $A=3$ nuclei (for a given $NN$ potential).

\subsection{Neutron magnetic form factor, $G_M^n$}

\begin{figure}[htpb]
\centering
\includegraphics[width = 0.65\textwidth]{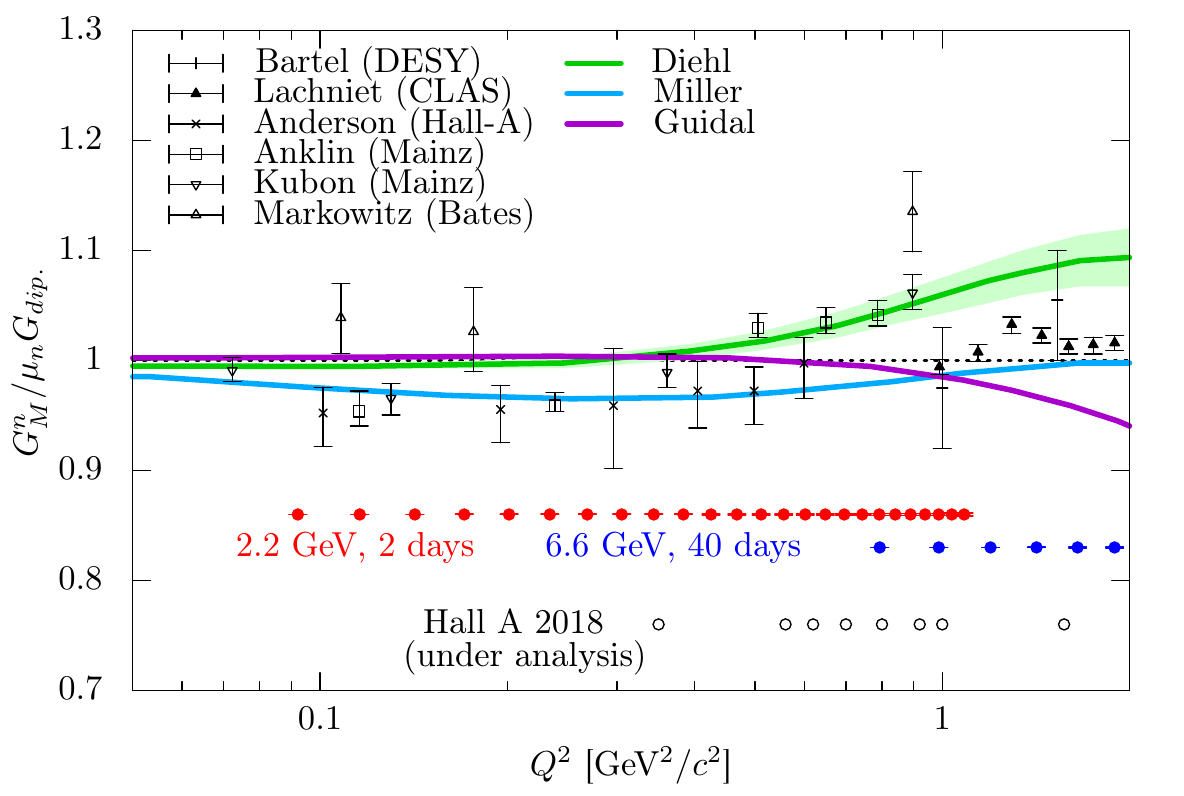}
\caption{\label{fig:GMn_reach} The anticipated kinematic
coverage and statistical uncertainty of the determination of $G_M^n$ compared to previous measurements~\cite{Bartel:1972xq,Lachniet_2009,Anderson:2006jp,Anklin:1994ae,Kubon:2001rj,Markowitz:1993hx} and theory~\cite{Miller:2002ig,Guidal:2004nd,Diehl:2004cx}.
}
\end{figure}

This experiment will measure $G_M^n$ at low ($Q^2<1$~[GeV/c]$^2$) and moderate-range $Q^2$. This measurement will use inclusive electron scattering from $^3$H and $^3$He targets at 2.2 and 6.6~GeV electron beam running. For the run period proposed in Table~\ref{requestedbeam}, we anticipate the statistics shown in Fig.~\ref{fig:GMn_reach}.

Fig.~\ref{fig:GMn_reach} shows the anticipated kinematic coverage and statistical uncertainty of the proposed experiment's extraction of $G_M^n$. In only one day of running on each target at 2.2~GeV, the proposed experiment will be able to thoroughly cover the kinematic region of $Q^2<1$~[GeV/c]$^2$, exactly where there are discrepancies between the measurement from Hall A~\cite{Anderson:2006jp} and previous measurements, and where GPD and cloudy-bag theory begin to diverge. In addition, running at 6.6~GeV will have $Q^2$ overlap with the 2.2~GeV data and will be able to extend the $Q^2$ coverage. Extracting $G_M^n$ from inclusive quasi-elastic data much above $Q^2=2$~[GeV/c]$^2$ will become difficult because of the increasing inelastic background. However, this region will be well-covered by CLAS12's run group B and at higher $Q^2$ by the Super-Big Bite program (both experiments using traditional scattering on deuterium), and is, therefore, not the focus of this proposal. Our measurements will complement these other experiments. 

\section{Proposed Measurement I: Quasi-elastics}~\label{QESection}
\subsection{Reaction mechanisms and event selection}

\subsection{$A(e,e'p)$ formalism}
Assuming factorization, the cross-section for electron-induced proton knockout from nuclei $A(e,e'p)$ can be written as:
\begin{equation}
\frac{d^6\sigma}{d\omega d E_{miss} d\Omega_ed\Omega_p}=K\sigma_{ep}S^D(E_{miss},P_{miss})
\end{equation}
where $\Omega_e$ and $\Omega_p$ are the electron and proton solid angles, respectively. $\sigma_{ep}$ is the cross-section for scattering an electron from a bound proton~\cite{DeForest:1983ahx}. $S^D(E_{miss},P_{miss}$) is the distorted spectral function. In the absence of final state interactions (FSI), $S$ is the nuclear spectral function that defines the probability to find
a nucleon in the nucleus with separation energy \emiss{} and momentum \pmiss. The missing energy and missing momentum are:
\begin{eqnarray}
E_{miss} &=& \omega - KE_p - KE_{A-1} \\
\vec{P}_{miss} &=& \vec{q} - \vec{P}_p
\end{eqnarray}
where $KE_p$ and $KE_{A-1}$ are the kinetic energies of the outgoing proton and residual nucleus. The momentum transfer $\vec q = \vec{P}_e - \vec{P'}_{e'}$, where $\vec{P}_e$ and $\vec{P'}_{e'}$ are the initial and scattered electron momenta. $\omega=E - E_0$ is the energy transfer, and $\vec{P}_p$ is the outgoing proton momentum. The kinematical factor, $K$, is:
\begin{equation}
K=\frac{E_p P_p}{(2\pi)^3}
\end{equation}

\subsubsection{Short range-correlated nucleon pairs}\label{section:otherprocesses}

\begin{figure}[htb]
\centering
\includegraphics[width=0.48\textwidth]{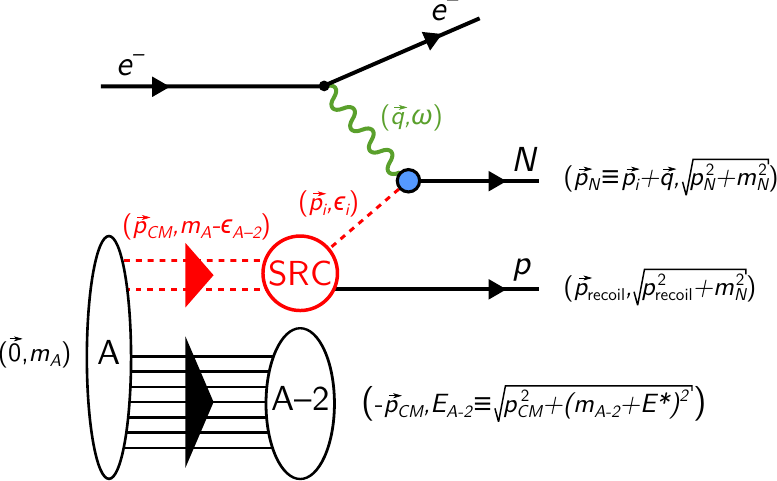}
\caption{The diagram of an incoming beam electron emitting a virtual photon and interacting with a nucleon in an SRC pair in the nucleus. The original nucleus is described by $A$, and the recoiling $A-2$ system is shown. The SRC pair yields the measured lead nucleon (proton in this proposal) along with the recoiling SRC-pair partner nucleon which may or may not be detected.}
\label{src_reaction} 
\end{figure}

Here we discuss the desirable kinematics in selecting events where the beam electron scatters from a nucleon in an SRC  pair, see Fig.~\ref{src_reaction}. 

\begin{figure}[htb]
\centering
\includegraphics[width=0.65\textwidth]{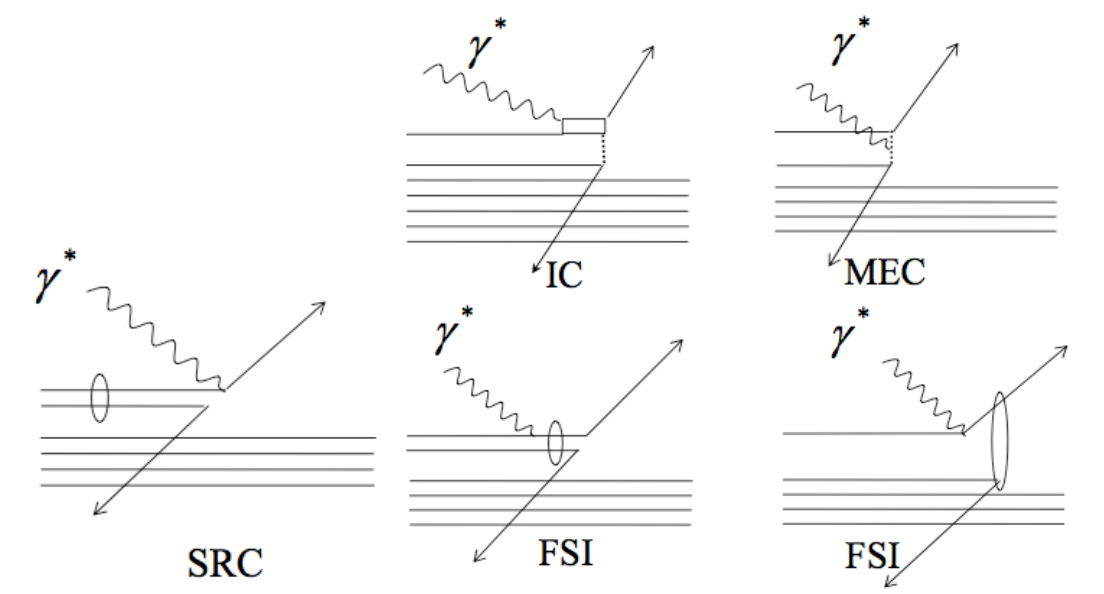}
\caption{Shown are the diagrams of the relevant processes in the kinematic region where we expect to find SRC nucleon pairs. The true SRC signal is shown on the bottom far left. The MEC and IC diagrams are shown on the top. The bottom center FSI diagram describes the FSI between the nucleons in the SRC pair. The bottom right describes the FSI of the nucleons with the resulting $A-2$ system.}
\label{competing_processes} 
\end{figure}

There are many competing processes in the QE  region that require us to fine tune our event selection criteria. These processes were mentioned in the Section~\ref{overview} and include Meson Exchange Currents (MEC), Isobar Configurations (IC), and Final State Interactions (FSI), see Fig.~\ref{competing_processes}.
The amplitude of the MEC diagram decreases faster than the SRC amplitude by a factor of $1/Q^2$. We can reduce the contributions from MEC by choosing $Q^2>2$~[GeV/c]$^2$. The contributions from both MEC and  IC processes are also suppressed at $x_B>1$. 

The bottom center FSI diagram describes the FSI between the nucleons in the SRC pair. In this case, the FSI between the nucleons in the pair conserves its  nucleonic composition and c.m. momentum, but changes $p_{rel}$. The bottom right describes the FSI of one of the nucleons with the resulting $A-2$ system. At  $Q^2\ge 2$ GeV$^2$  the struck proton  is ejected with enough momentum ($P_p>1$~GeV/c) so that we can use the Glauber approximation to describe  FSI. In addition, we can minimize the effects of FSI  by selecting the angle between the recoil $A-1$ system and $\vec q$, $\theta_{rq}<40\degree$.
Table~\ref{generalSRCcuts} shows the previously-used SRC event selection criteria.

\begin{table}[htb!]
\caption{\label{generalSRCcuts} Previously used \eep{} and \eepN{} SRC selection criteria.  $\theta_{pq}$ is the angle between the knocked out proton and $\vec{q}$.   $m_{miss}^2=[(\vec{q},\omega)+(\vec{0}),m_d)-(\vec(P),E_p)]^2$ is the missing mass of the reaction assuming scattering off a stationary nucleon pair. $m_d$ is the deuteron mass and $E_p$ is the proton final energy from $E_p=\sqrt{\vec{P}^2+m_p^2}$.}
\centering
\begin{tabular}{ | C{5cm}  | }
 \hline
 $x_B >1.4$ \\
 \hline
 $300<| \vec P_{miss} |< 1000$~MeV/c \\
\hline
$\theta_{pq} < 25^{\circ}$ \\
\hline
$\theta_{rq} < 40^{\circ}$ \\
\hline
$0.6 <| \vec{p}| / | \vec{q} |< 0.96$ \\
\hline
$m_{miss} < 1.1$~GeV \\
\hline
  \end{tabular}
\end{table} 

The cuts on $\theta_{pq} < 25^{\circ}$ and $0.6 <| \vec{p}| / | \vec{q} |< 0.96$ select the "leading" proton, i.e., the proton that absorbed the virtual photon.  The $x_B>1.2$ or 1.4 cut and  the $Q^2\ge 2$ GeV$^2$ cut reduce the effects of MEC and IC.  , The $m_{miss} < 1.1$~GeV cut eliminates events with an undetected pion (typically from IC or other resonant processes). The $\theta_{rq}$ cut reduces the effects of FSI.  The \pmiss{} cut selects protons from SRC pairs.

\subsection{SRC observables}

We propose to measure  $A(e,e'p)$, and $A(e,e'pN)$ reactions on $^3$He and $^3$H using the CLAS12 spectrometer. We will use hydrogen elastic data to calibrate and normalize our data. We will also measure $d(e,e'p)$  to further understand contributions from FSIs and make comparisons. (The $A(e,e')$ reaction will be used to extract the ratio of $\sigma_n/\sigma_p$ for obtaining the neutron magnetic form factor, $G_M^n$.) 

For studying few-body nuclear structure and short range correlations, we will measure absolute cross sections for each target over $0 \le \pmiss \le 1$  1~GeV/c.  We will measure the semi-inclusive  $A(e,e'p)$ cross sections to extract nuclear momentum distribution for the $A=3$  nuclei to compare to cross-section and momentum-distribution calculations. The isospin asymmetry between the $A=3$ targets and the large kinematical coverage of the CLAS12 detector will enable us to study the reaction channels, evaluate non-QE contributions, and make strong comparisons with theoretical predictions.

We will also measure \eepN{} cross sections and ratios, in order to measure the characteristics of the SRC pairs.  These include their isospin/spin composition ($pp$, $nn$ or $pn$ pairs and either spin 1 or spin 0), their center-of-mass momentum distributions, and their relative momentum distributions.  We also want to understand how their spin/isospin composition changes with missing momentum, in order to understand the tensor to scalar transition previously seen around 600 MeV/c \cite{Schmidt:2020kcl}.

\subsection{Experimental setup and kinematical coverage}

\begin{figure}[htb]
\centering
\includegraphics[width= 0.45\textwidth]{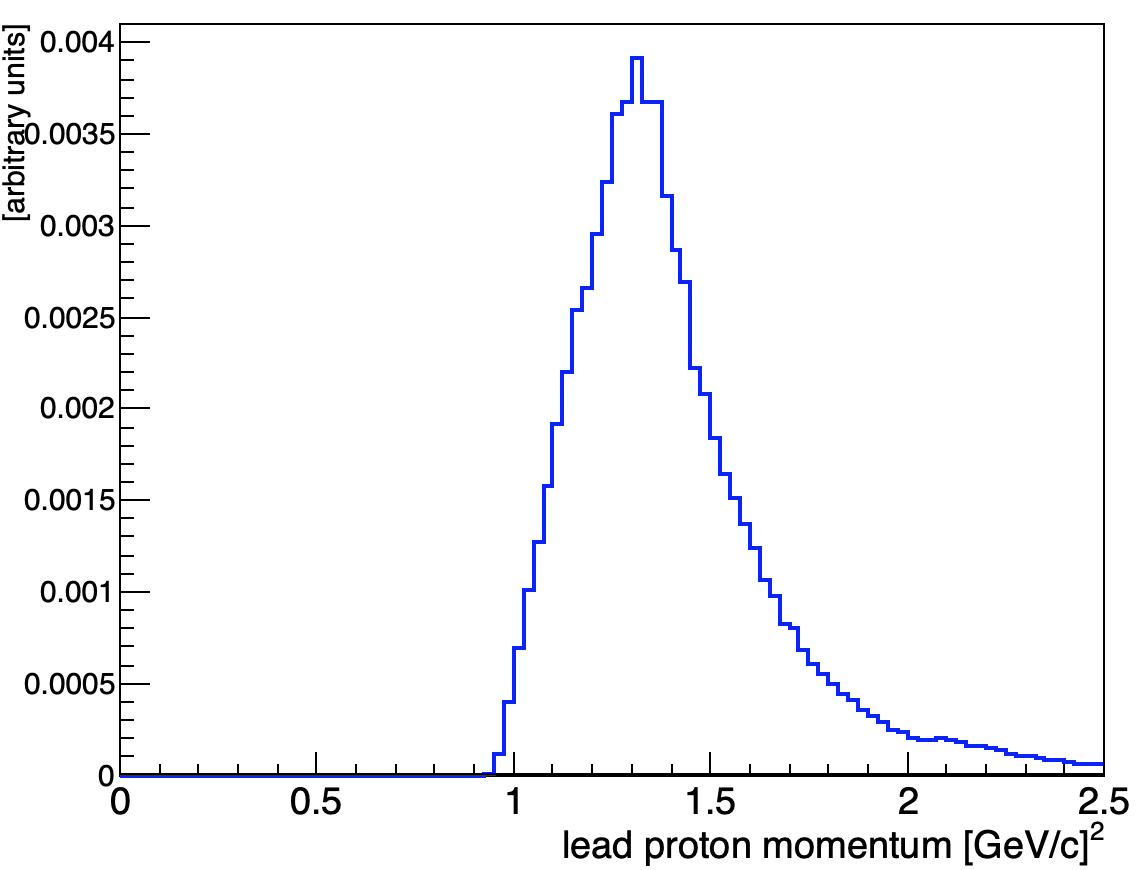}
\includegraphics[width= 0.45\textwidth]{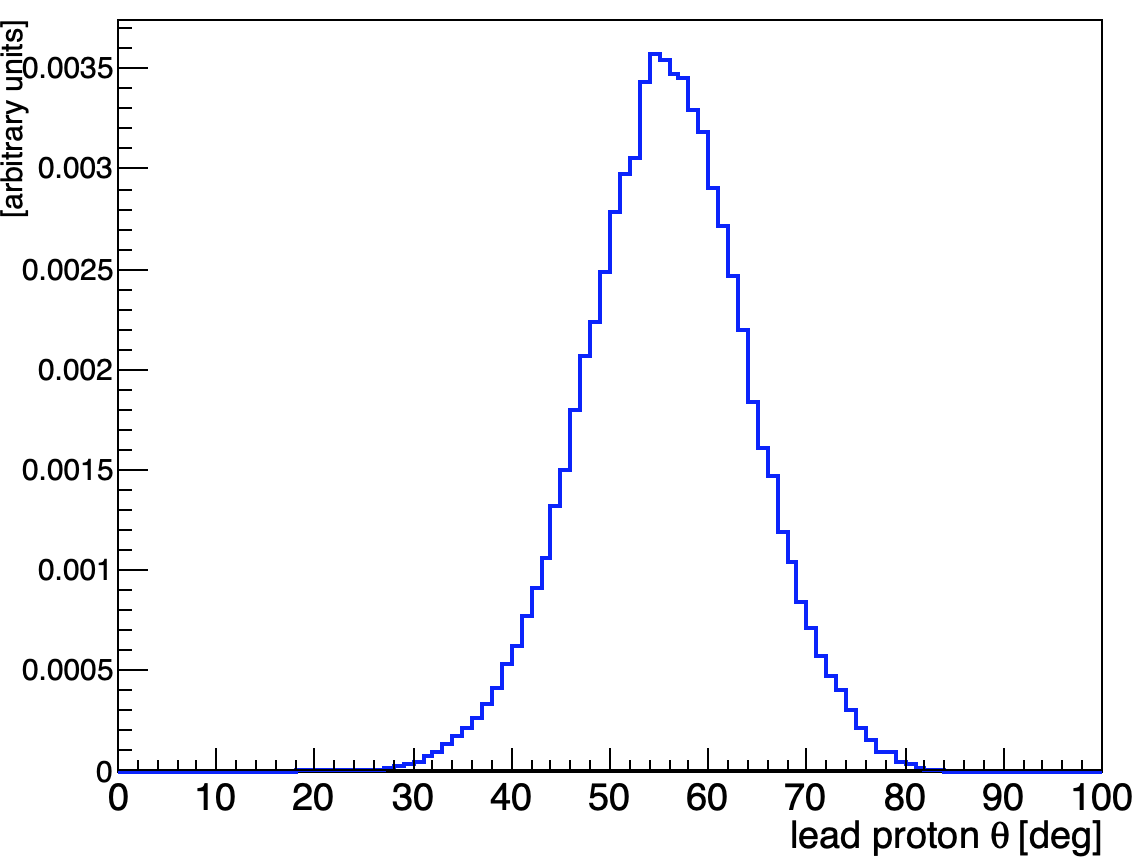}
\caption{Left: The  momentum of the leading proton. Right: The angle of the leading. These distributions include the lead proton cuts described in Table~\ref{generalSRCcuts} with approximate electron fiducial cuts.}
\label{fig:leadPspectrum}
\end{figure}

We will use the large-acceptance and open (i.e., electron only) trigger of the CLAS12 detector to measure inclusive and semi-inclusive hard scattering from $^3$H, $^3$He, and $^2$H targets. While our proposal focuses on  events where the leading nucleon is a proton, we will also be able to detect leading neutrons in the EC, as well as spectator correlated neutrons in the central neutron detector and BAND. The expected lead proton momentum and angle are shown in Fig.~\ref{fig:leadPspectrum}.

\begin{figure}[htb]
\centering
\begin{minipage}{0.45\textwidth}
\includegraphics[width=\textwidth]{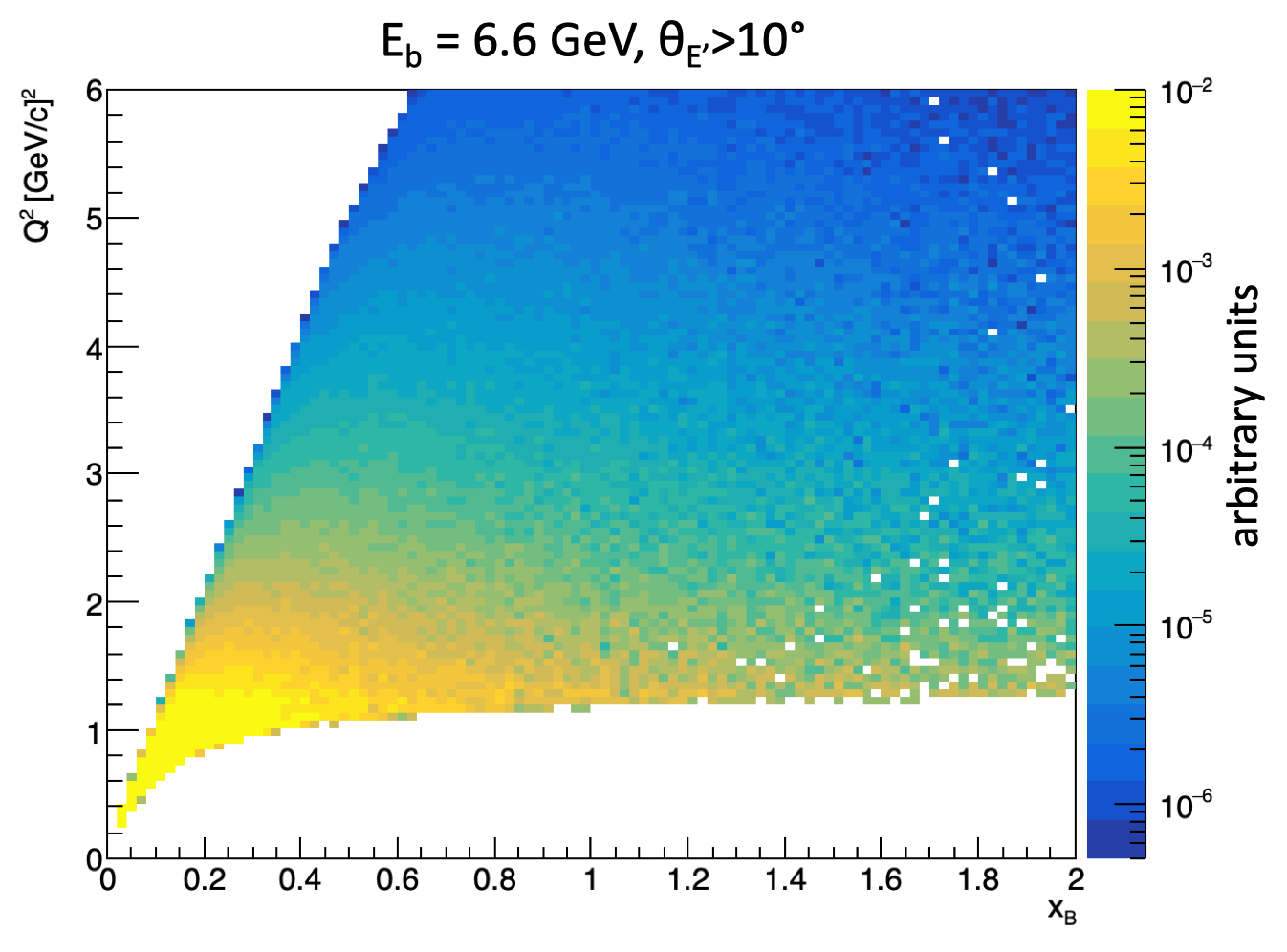}
\end{minipage}\begin{minipage}{0.45\textwidth}
\includegraphics[width=\textwidth]{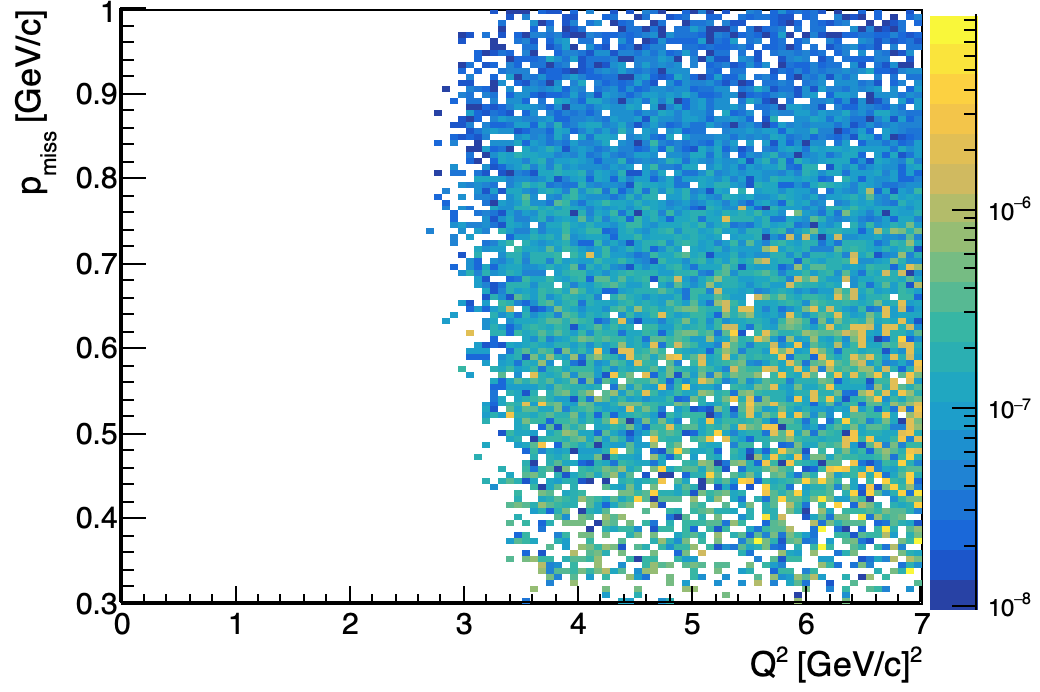}
\end{minipage}
\caption{Left: The general phase space acceptance for electrons in the CLAS12 detector at minimum angles greater than 10\degree for an electron beam of 6.6~GeV. Right: The distribution of minimum accessible $Q^2$ for an incident 6.6~GeV electron beam for the selection criterion listed in Table~\ref{generalSRCcuts}. This distribution is generated using the GCF AV18 interaction and the $z-$axis units are arbitrary.}
\label{fig:clas12_6p6}
\end{figure}

The majority of our \eep{} and \eepN{}  events will be measured with an incident beam energy of 6.6~GeV. We used a simple phase-space generator  to calculate the electron scattering acceptance. Rough CLAS12 fiducial cuts (determined by F.X.~Girod from the standard GEMC CLAS12 geometry) were applied to the electron spectra. CLAS12 has a large kinematical acceptance as shown on the left in Fig.~\ref{fig:clas12_6p6}. In this proposal, we conservatively estimate rates and acceptances for scattered electrons at $\theta_e\ge  10\degree$.   

The event selection cuts of Table~\ref{generalSRCcuts} dramatically increases the minimum accessible $Q^2$ (see Fig.~\ref{fig:clas12_6p6}right).  

\begin{figure}[htpb]
\centering
\includegraphics[width=0.5\textwidth]{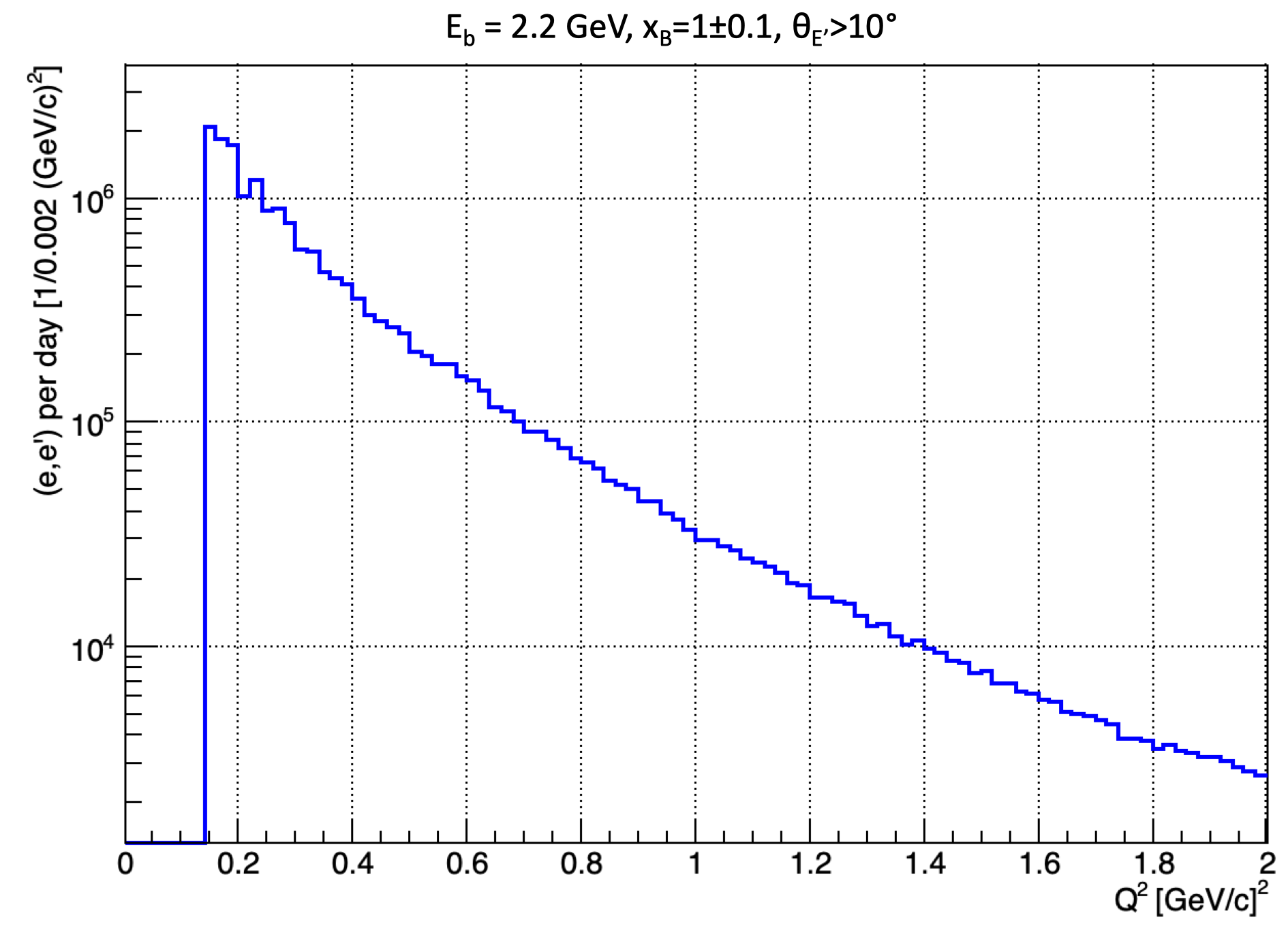}
\caption{The inclusive rate is shown for a 2.2~GeV electron beam as a function of accessible $Q^2$. The minimum angle here is 10$\degree$. The $Q^2$ range is truncated at 2~GeV$^2$/c$^2$ beyond which we will not extract $G_M^n$.}
\label{fig:clas12_inclusive}
\end{figure}

Fig.~\ref{fig:clas12_inclusive} shows the expected inclusive electron scattering rate as a function of $Q^2$ for the 2.2 GeV data.

The 2.2~GeV inclusive rates shown in Fig.~\ref{fig:clas12_inclusive} were compared to $^3$He inclusive rates in the e2a experiment with the same beam energy where the minimum electron scattering angle was 20\degree. In CLAS12, the minimum scattering angle in Fig.~\ref{fig:clas12_inclusive} is cut off at 10$\degree$ and significantly extends the reach to low $Q^2$, a crucial region needed for precisely determining $G_M^n$. 

\subsection{Rate Estimation and beam time request}
 The rate estimation uses the measured  $A(e,e'N)$ event rates on carbon from the CLAS6 EG2 experiment~\cite{Hen2014nza}. The EG2 experiment ran for 25 PAC days at an incident beam energy of 5~GeV on solid targets with a deuterium target cell in the beam, simultaneously. We doubled the rate to account for our single target cell. The effective usable gas-target luminosity of our proposed experiment is approximately the same as the luminosity of EG2 ($10^{34}$~cm$^{-2}$s$^{-1}$). Therefore, the carbon rates are scaled by a factor of two to account for the increased solid angle acceptance, Mott cross-section (5 to 6.6 GeV beam energy), and recoil detection. Assuming $A^{-1/3}$ scaling (corresponding to the probability that a knocked out nucleon in a nucleus $A$ escapes the nucleus where nucleons in carbon have a 40\% chance to escape versus the nucleons in $A=3$ that have a 70\% chance to escape), the rates are increased by a factor of 1.6 from the $A=12$ carbon nucleus to the $A=3$ nuclei. 

As a check, we used the $A=3$ Cracow calculations~\cite{Golak2005ElectronAP, caracsco} (used in the Hall A tritium study in Ref.~\cite{Cruz-Torres:2020uke}) to generate the total cross section for discrete kinematic points.  We compared this cross section  with the scaled measured C\eep{} yields.  The Cracow calculations give about  $1/2-1/3$ as many events as the carbon data. This is roughly consistent with ratio of the AV18 extracted contact terms of carbon to the $^3$H and $^3$He nuclei which yields a ratio of approximately $7/18$~\cite{cruztorres2019scale_supp}. Therefore, we applied a conservative factor of $1/3$ scaling to the scaled C\eep statistics. 

The number of beam days per target is shown in Table~\ref{requestedbeam}. We request 20 days for high statistics running on each $A=3$ target at 6.6~GeV. This should give us about 6000 \eepN{} events each with a leading proton  and a recoil proton or neutron, see  Table~\ref{expectedcounts}. The \eepn{} event rate for \trit{} and \het{} should be similar at large \pmiss.  Deuterium has approximately half the number of SRC pairs compared to $^3$H (and therefore, half of the statistics at high missing momentum for the same run time). We require deuterium to constrain FSIs in order to better calculate reactions on $A=3$ nuclei. Deuterium having 25\% statistics (compared to $^3$H) in the high missing momentum regime is sufficient for this experiment. Therefore, we request 10 days of running on deuterium. We will use hydrogen to optimize our calibration and measure the absolute cross section for efficiency studies and systematic effects. Table~\ref{requestedbeam} includes 1 day (3 shifts or 0.5 PAC days) of overhead for each target change. 

\begin{table}[htb!]
\caption{\label{requestedbeam}Requested beam time per target, including calibration time and target change overhead.}
\centering
\begin{tabular}{ | p{5cm} | C{1cm} | C{1cm} | C{1cm} | C{1cm} | C{1cm} | }
 \hline
 \textbf{Target:} & \textbf{$^1$H} & \textbf{$^2$H} & \textbf{$^3$He} & \textbf{$^{3}$H} & \textbf{Total}\\ 
  \hline
 Measurement Days (6.6 GeV) & 1 & 10 & 20 & 20 & \textbf{51}\\ 
  \hline
  \multicolumn{5}{|l|}{Calibration (inbending field) } &  \textbf{1}\\ 
 \hline
 \multicolumn{5}{|l|}{Target Changes } &  \textbf{2}\\ 
  \hline
    \multicolumn{5}{|r}{\textbf{Total at 6.6 GeV: }} & \textbf{54}\\ 
 \hline
 Measurement Days (2.2 GeV) & 0.5 & 0 & 1 & 1 & \textbf{2.5}\\ 
   \hline
  \multicolumn{5}{|l|}{Calibration (outbending field)} &  \textbf{1}\\ 
 \hline
 \multicolumn{5}{|l|}{Target Changes } &  \textbf{2}\\ 
  \hline
    \multicolumn{5}{|r}{\textbf{Total at 2.2 GeV: }} & \textbf{5.5}\\ 
 \hline
   \multicolumn{5}{|r}{\textbf{Total beam time requested: }} & \textbf{59.5}\\ 
 \hline
  \end{tabular}
\end{table} 
The field will be in the electron out-bending configuration for the 2.2~GeV beam running and will be in the electron in-bending configuration for the 6.6~GeV beam energy run. We include one pass change from 2.2~GeV beam energy to 6.6~GeV beam energy that should take half of a shift. 

\begin{table}[htb!]
\caption{\label{expectedcounts} Expected number of counts for 2N knockout reactions for 20 beam days on each target $^3$He and $^3$H at 6.6 GeV beam energy. The reaction notation is that the first nucleon is the ``leading" nucleon (i.e., a high-momentum nucleon that is emitted largely in the momentum transfer direction), and other nucleons are the recoil nucleons. We consider cases only where the leading nucleon is a proton.}
\centering
\begin{tabular}{ | C{3cm}  | C{18mm} | C{18mm} | C{18mm} | C{18mm} | }
 \hline
 Reaction & $(e,e'pp)$ & $(e,e'pn)$ \\
 \hline
 \# events (6.6 GeV) & 8k & 6k \\
\hline
  \end{tabular}
\end{table} 

For the $A(e,e'p)$ reaction specifically, we anticipate the statistics for the requested beam time in the high missing momentum regime as shown in Figs.~\ref{fig:A3-proj} and \ref{fig:sum-proj}, which will enable us to discern between various theoretical models. This experiment uniquely accesses this regime with high statistics on these calculable nuclei.

As detailed in the next Section \ref{TargetSection}, for our expected luminosity on \trit{} of $2\times10^{34}$~cm$^{-2}$s$^{-1}$, we intend to run at a requested beam current of 110~nA. This configuration will not require rastered beam. 

\input{Sections/target-design}

\subsection{Conclusion}
We conclude that a tritium target similar to the system developed for use in Hall A could be similarly employed in Hall B. While an exhaust system would need to be developed, there are many other aspects of the proposed Hall B system that would make the design and fabrication more simple. The proposed target cell would be installed on a dedicated insertion cart for the duration of the tritium run. Therefore, no motion system is required. Further, with the use of a dedicated cryo-cooler and lower beam current the cell and heat sink design are also simplified. Based on a similar analysis performed for the HATT, a release, in a controlled fashion through the stack or through the truck ramp, of the full load of tritium contained in the cell is not expected to pose a significant risk to personnel on site or to the public.

Because the Hall A target and the proposed Hall B target are very similar, the budget for each system is also expected to be similar. Some expenses that were incurred in the Hall A project, such as the tritium exposure study of aluminum 7075 will not have to be repeated.

\section{Relation to other approved 12 GeV measurements}

There is no experiment to date that probes both $^3$He and $^3$H across the full quasi-elastic kinematical regime. While the Hall A tritium experiments showed that we can learn much from these isospin asymmetric targets, our fundamental understanding of the $NN$ wave function will only be fully constrained from studies on both targets with a more thorough evaluation of the non-QE reaction mechanisms. Including deuterium will improve our understanding of FSIs in the limit of a two-body system. These studies will be naturally accessible from the large acceptance of the CLAS12 detector and will yield new measurements to higher \pmiss{} where different momentum distribution models can be tested. Our measurements will support the interpretation of the Hall A \trit{} spectrometer measurements and will improve our interpretation of future experiments on heavier nuclei such as the study of SRCs using CLAS12 in Run Group M. 

Additionally, our experiment will also measure $G_M^n$ by comparing the inclusive scattering cross sections from $^3$He and $^3$H, covering the crucial low $Q^2$ regime with different systematics than measurements from deuterium. These measurements will be a significant improvement on the previous measurements of $G_M^n$ at low $Q^2$ and will support the overall understanding of $G_M^n$ which will be an important interpretation to the recent Run Group B measurement and future SBS measurement. 


\bibliographystyle{unsrt}
\bibliography{biblio.bib}

\end{document}

%% file: Sections/target-design.tex
\section{Target design}\label{TargetSection}

A new gas target system is proposed for this measurement. While a detailed conceptual design is not presented here, we propose that such a system should build on the experiences and lessons learned from the Hall A Tritium Target (HATT) which is described in detail in reference \cite{HATTrpt}. While this system would be unique, the same rigor that was applied to the Hall A system would also be applied in Hall B. The safety systems and subsystems for the proposed target are necessarily complex and can only be summarized in this proposal. This includes numerous engineered and administrative controls, only some of which are listed below.
\begin{itemize}
    \item A minimum of three layers of containment and or confinement shall be employed at all times. This includes operations, installation/removal, shipping and handling, and storage.
    \item The cell shall be constructed in compliance (and indeed in excess of compliance) with JLAB pressure safety requirements, SRTE safety basis requirements, and applicable ASME Codes, namely ASME B31.3 and ASME BPVC VIII D1 and D2. Design safety factors for the cell shall exceed 10. This shall be verified by through destructive testing.
    \item Strict access controls shall be required for the Hall while the cell is installed. These include specific training, locked badge access to the Hall including truck ramp access, and procedures for entrance/exit of the Hall. These controls are partly to ensure that the Hall will be the third layer of confinement while the cell is installed.
    \item An extensive review process shall be employed, specifically:
    \begin{itemize}
        \item Technical and Peer reviews as required my EHSQ 6151 and supplement.
        \item Review by Savannah River Tritium Enterprises and DOE-NNSA.
        \item Multiple reviews by JLAB and outside Subject Matter Experts (SME) as part of the formal ERR process addressing all aspects of the system.
    \end{itemize}
    \item Full Failure Mode Effects and Criticality Analysis (FMECA) shall be performed and reviewed by SMEs. All failure modes shall be addressed including complete failure of the containment/confinement system.
    \item Examinations of materials, completed components, welds, mechanical fabrications, etc. shall be performed by qualified personnel in accordance with approved procedures.
    \item Inspections verifying all appropriate examinations have been performed and documented by qualified personnel using correct procedures and calibrated/certified equipment.
    \item Specific training and additional qualifications shall be required for all personnel accessing the Hall and performing any fabrication function.
    \item A thorough review and site inspection by EHS and Physics Div SMEs ensuring all applicable safety systems are installed and are operating correctly shall be performed.
\end{itemize}

The proposed system would employ three sealed gas cells filled with $^{2}H_2$, $^{3}H_2$, and $^{3}He$. Given the limited space in CLAS12, a motion system is not possible, and the cells would need to be installed separately, marking three distinct run periods. The Hall B tritium target is expected to incorporate the same major components as the Hall A system which are listed below.

\begin{itemize}
	\item Target Cell
	\item Exhaust system including stack
	\item Containment/confinement system including the scattering chamber and Hall B under strict access controls.
	\item Cryogenic cooling system
\end{itemize}

Some conceptual design work has been performed and is shown in the subsections below. Rate estimates reported elsewhere in this proposal are based, at least in part, on the thicknesses, materials, and geometries presented in this concept.

\subsection{Tritium Containment and Confinement}
The primary method for ensuring safe operations with tritium is to establish multi-layer containment and or confinement at all times. The proposed system would rely on a series of engineering and administrative controls to provide at least three layers of tritium confinement and/or containment during all phases of operation. Confinement as defined here would limit a possible tritium release to a controlled region were it would be collected and stacked (exhausted to the environment) in a safe manner. A summary of the three layers of containment/confinement are shown in the table below for each operational condition (configuration).

\vspace{3 mm}
\begin{tabular}{|p{1.2in}|p{0.5in}|p{1.5in}|p{1.5in}|}
	\hline 
	Configuration & Layer 1 & Layer 2  & Layer 3 \\ 
	\hline 
	Installation/Removal & Cell & Handling Hut and Scattering Chamber & Hall B \\ 
	\hline 
	Shipping/Storage & Cell & Inner Containment Vessel &  Outer Containment Vessel \\ 
	\hline 
	Beam Operations & Cell & Scattering Chamber & Hall B  \\ 
	\hline 
\end{tabular} 
\vspace{3 mm}

It is important to note that during beam operations, the Hall and scattering chamber must each be considered as one layer of the confinement system. This has implications for the design of the scattering chamber. It also requires that the exhaust system and access controls are designed to ensure that the Hall and chamber can indeed be considered layers of confinement.

\subsection{Target Cell}
\begin{figure}[htb]
\centering
\includegraphics[width=0.8\textwidth]{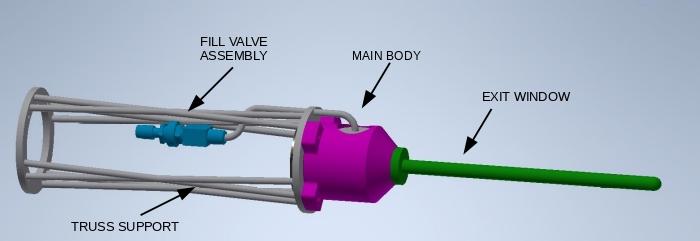}
\caption{External side view of the conceptual design of the target cell as seen from the beam right side.}
\label{T2-cell} 
\end{figure}

\begin{figure}[htb]
\centering
\includegraphics[width=0.8\textwidth]{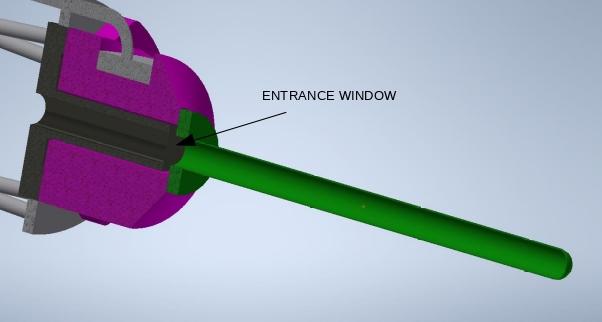}
\caption{Section view of the conceptual design of the target cell.}
\label{Cell-section} 
\end{figure}

\begin{figure}[htb]
\centering
\includegraphics[width=0.8\textwidth]{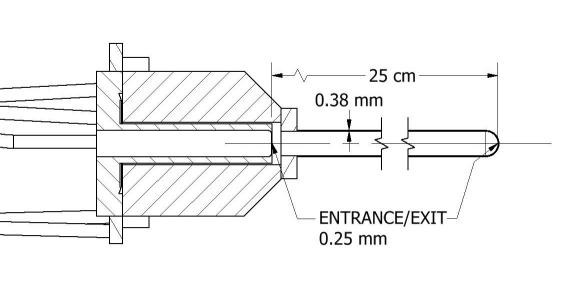}
\caption{Detail side section view of the target cell with thickness and length. Note that the active region of the cell is 25 mm long.}
\label{Cell-detail} 
\end{figure}

A conceptual model of the cell is shown in Figures~\ref{T2-cell},~\ref{Cell-section}, and~\ref{Cell-detail}. With the exception of the fill valve assembly, the cell is fabricated from ASTM B209 7075-T651 aluminum. This material has many distinct advantages, primarily, being nearly twice the strength and hardness of more common alloys (e.g. 6061). It has also undergone Jefferson Lab sponsored testing at Savannah River National Laboratory confirming suitability for tritium service at our operating conditions \cite{Duncan2017}. The proposed cell for the Hall B target allows full azimuthal angle acceptance and backward polar angle acceptance, with minimal loss of target length, to $120\degree$. The azimuthal symmetry also greatly simplifies the design of the target cell making it much easier to fabricate than the HATT cell. The target cell is expected to be 12.7 mm in diameter and 25 cm long with a fill pressure of about 200 psi. Thus, the total amount of tritium would be about 1200 Ci. The thickness of the cell wall is 0.4 mm with the exception of the beam entrance and exit which are expected to be 0.25 mm. While these thicknesses are not optimal when considering the physics, they do provide a suitable level of safety both during beam operations and during the filling of the cell off site. Similar sealed gas cells were used in Hall A for the Tritium Family of experiments and performed at $22.5 \mu A$ with acceptable density reduction \cite{Santiesteban2019a}. Filling of the tritium cell is expected to be performed at Savannah River Site where overpressure protection requirements are substantially higher than the fill/operating pressure of the cell.

Hydrogen is known to permeate through most materials. A model was developed for the expected permeation of tritium from the HATT cell \cite{Permcalc2017}. The expected operational loss of tritium from the cell is less than 0.8 Ci per year due mostly to permeation through the thin cell walls. This is similar to the loss observed in Hall A during operations. This loss, although small, would be collected by the pumping system and stacked.

The temperature of the cell wall should not exceed 170K for extended periods of time. This is a design requirement based on previous studies of hydrogen embrittlement in aluminum with an impinging electron beam \cite{Flower1977a}. In Hall A, the cooling system was supplied by 15K helium from the ESR. For practical reasons, a dedicated cooling system (similar to that of the Hall D cryogenic target) should be used for a Hall B target. A stand alone pulse-tube refrigerator system such as the CryoMech PT410 with a cold finger would simplify the design and provide operational reliability. The beam current necessary to complete the measurements is much less than $ 1 \mu A$. The heat generated in the cell with this current would be less than 1 W, with the majority being generated in the entrance and exit windows. A thermal model of the cell exit window (the component most affected by the heat load) has been developed assuming following conditions

\begin{itemize}
    \item Beam current $1 \, \mu A$
    \item Beam spot size of 0.250 mm
    \item Cold sink operating temperature 40K
    \item Cold finger length 3 m
    \item Cold finger cross section $1 \, cm^2$
    \item Model is run in steady state only
\end{itemize}

\begin{figure}[htb]
\centering
\includegraphics[width=0.8\textwidth]{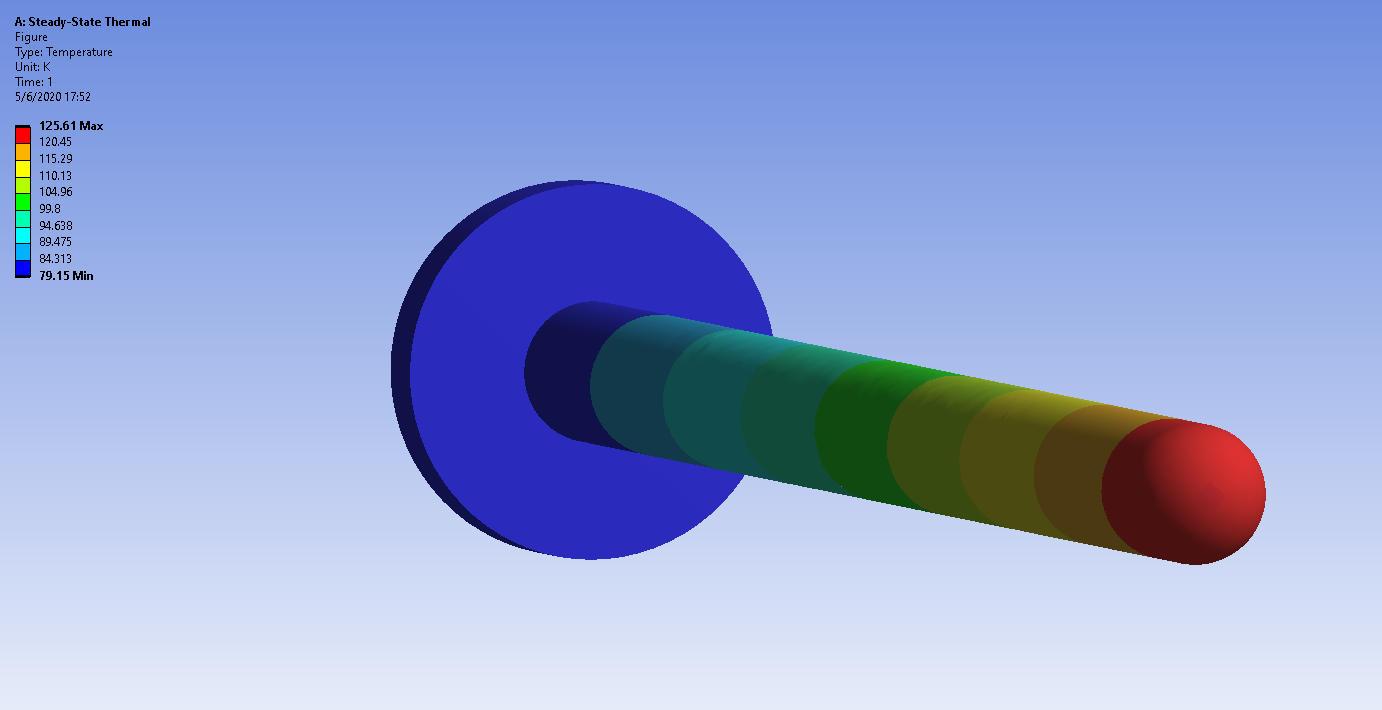}
\caption{Temperature profile of the aluminum exit window under the conditions listed above. The maximum temperature (red) is ~126K and minimum (blue) is ~80K}
\label{Cell-temp} 
\end{figure}

\begin{figure}[htb]
\centering
\includegraphics[width=0.8\textwidth]{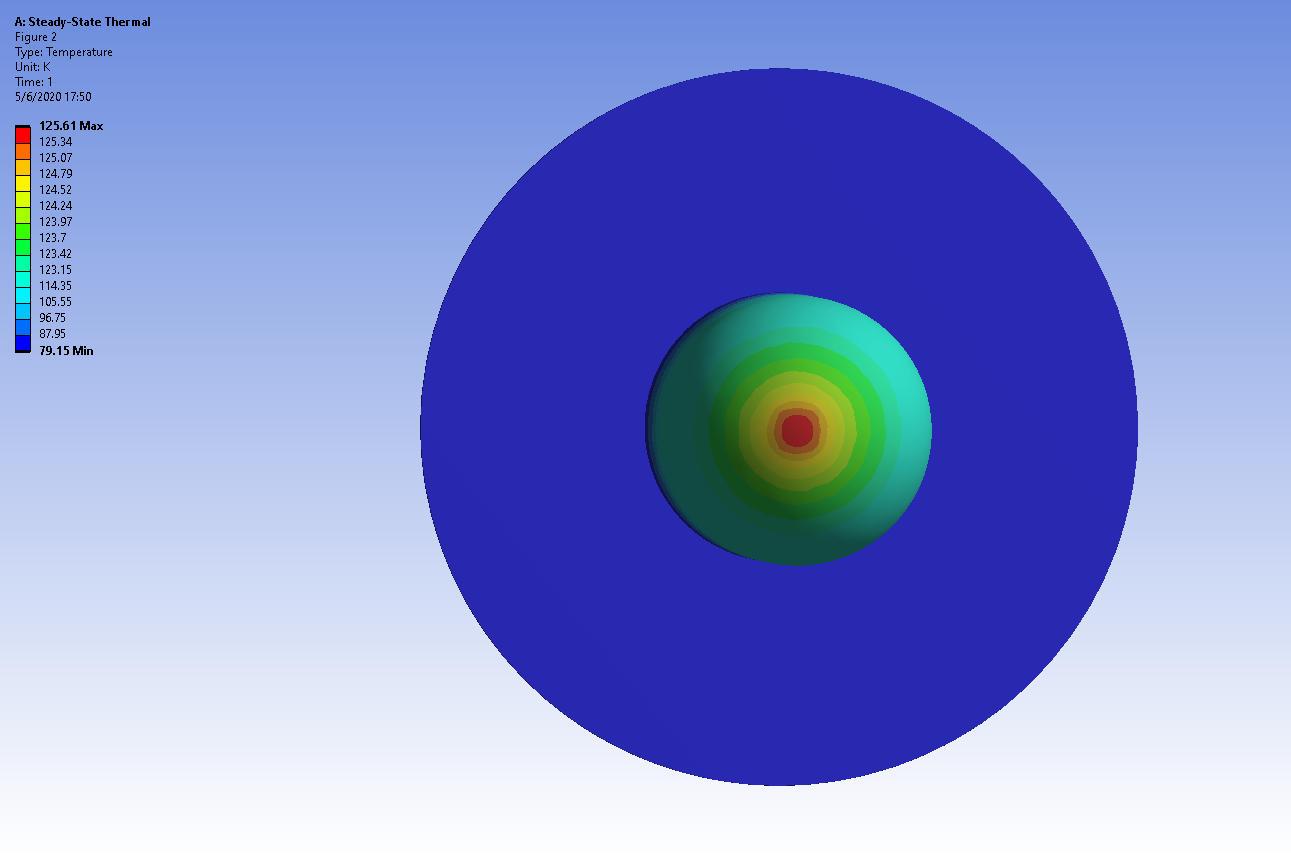}
\caption{Temperature profile of the exit window tip under the beam conditions listed above. Note that the beam passes through the dark red area in a uniform 0.250 mm spot size and that the temperature scale in this figure has been logarithmically inflated near the tip.}
\label{Cell-temp-tip} 
\end{figure}

The model was developed using an ANSYS steady state thermal analysis. The results of this simulation are shown in Figures \ref{Cell-temp} and \ref{Cell-temp-tip} 

\subsection{Scattering Chamber and Vacuum System}
\begin{figure}[htb]
\centering
\includegraphics[width=0.8\textwidth]{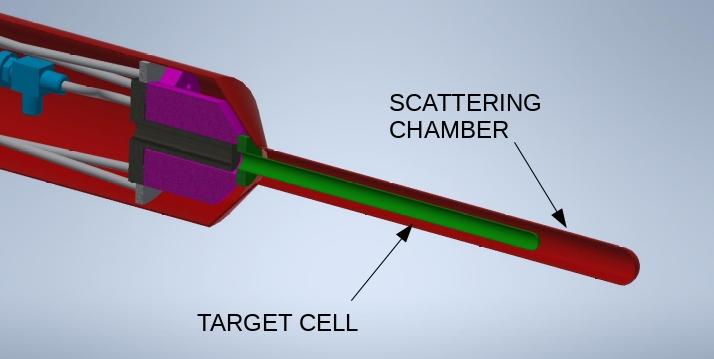}
\caption{Section view of the target cell/chamber assembly }
\label{total-assy} 
\end{figure}

\begin{figure}[htb]
\centering
\includegraphics[width=0.8\textwidth]{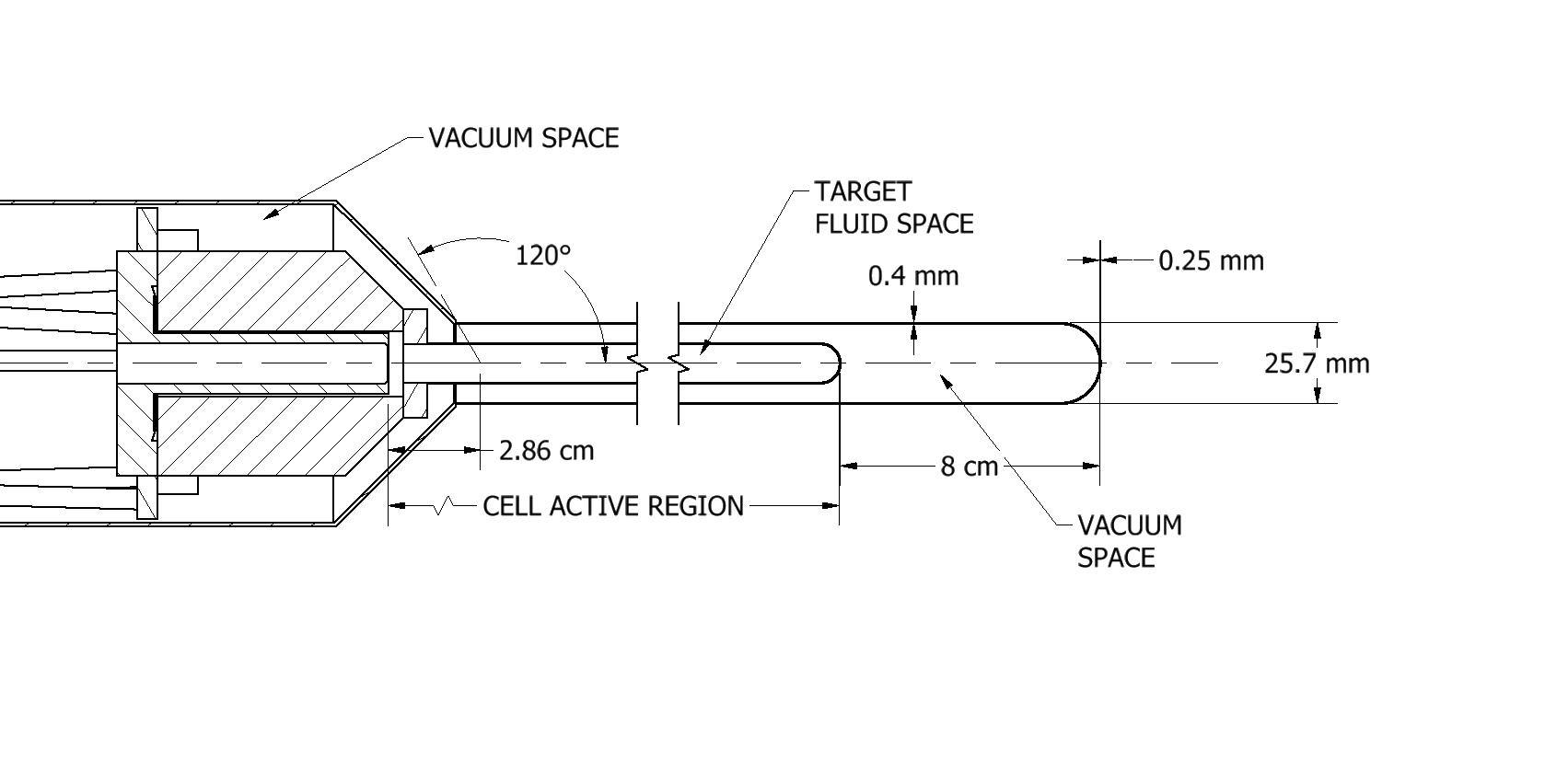}
\caption{Detail section view cell/chamber assembly with some dimensional detail. Note that for $120\degree$ the effective length of the target is shortened by less than 3 cm.}
\label{assy-dwg} 
\end{figure}

Because the scattering chamber must be a layer of tritium confinement, it cannot be fabricated from materials that are significantly permeable to tritium. Thus, the foam chamber typically used in Hall B will have to be replaced. We propose a chamber fabricated from aluminum (ASTM B209 7075-T651) with geometry as shown in Figure \ref{total-assy}. In this design, the part of the chamber surrounding the cell is cylindrical with a hemispherical head and has a diameter of 25 mm. Some details of the chamber geometry are shown in Figure \ref{assy-dwg}. Calculations (in compliance with ASME Boiler and Pressure Vessel Code Sec VIII Div 1 and Div 2) show that a wall thickness of 0.4 mm will exceed a safety factor of 2 for buckling from the external pressure. The beam exit of the chamber (tip of the hemisphere) shall be thinned to 0.25 mm. The chamber must also be isolated from the upstream beam line via a 0.2 mm thick beryllium window. Additionally, the chamber volume must be large enough to contain a gas cell failure and still maintain a sub-atmospheric pressure. The vacuum in the chamber would be maintained by a series of mechanical and turbo pumps which would exhaust to the stack.

\subsection{Luminosities}
It is important to consider that the aluminum walls of the target cell and scattering chamber in the proposed design are relatively much thicker than the nominal Hall B cryotarget configuration. The reasons for this have been discussed in the previous sections. A summary of the assumed thickness and luminosity for the tritium cell, both fluid and metallic components, is shown in Table~\ref{tab:tlumi}. The fluids in the  $^{2}H_2$ and $^{3}He$ cells are expected to have fluid densities of 125\% and 150\% of the tritium cell. The entrance and exit windows (Al windows) for the cell and chamber have been combined into one thickness.

\begin{table}[htb]
    \centering
    \begin{tabular}{c|c|c|c|c}
        Material & Tritium & Al Windows & Be Window & Total \\
        \hline
        Length(g/cm$^2$) & 0.085 & 0.21  & 0.037 & 0.33\\
        \hline
        Luminosity & $3.54\times 10^{34}$ & $8.42\times 10^{34}$ & $1.54\times 10^{34}$ & $1.35\times 10^{35}$\\
    \end{tabular}
    \caption{Assumed density and luminosity for each component. Note that the maximum luminosity of CLAS12 is $1.35\times 10^{35}$ nucleon/cm$^2$/s}
    \label{tab:tlumi}
\end{table}

\subsection{Exhaust System and Stack}
A dedicated exhaust system and stack shall be constructed to remove tritium from Hall B similar to the system developed for Hall A. Calculations for tritium release resulting from a catastrophic cell failure were performed using HotSpot \cite{Homan2015a}. This is a DOE approved collection of atmospheric dispersion models which are designed for near-surface releases, short-range (less than 10 km) dispersion, and short-term (less than 24 hours) release durations in unobstructed terrain and simple meteorological conditions. These calculations, summarized in \cite{HATTrpt}, indicate the exhaust stack must be at least 20~m above grade at the site boundary. (Note the the position of the Hall B stack would be less than 15~m from the current Hall A stack.) This ensures that any release will not cause an undue exposure to the public. The exhaust system shall be driven by a fan which pulls air through Hall B (maintaining a slight negative pressure) into the stack via one of the smoke removal ports. Thus, the exhaust system serves two purposes: tritium removal and smoke removal. This  system must also stack the exhaust from the vacuum pumps connected to the scattering chamber and downstream beam line. These subsystems are necessary to ensure at least three layers of containment or confinement as discussed previously.

\subsection{Transportation and Storage}
The HATT cell was filled at Savannah River Site and shipped to Jefferson Lab in the Bulk Tritium Shipping Package (BTSP) as a miscellaneous tritium vessel (MTV). The same mechanism is expected to be employed for filling and shipping a similar cell for the Hall B target. An expert team from SRS traveled to Jefferson Lab to assist in the unpackaging and packaging of the cell to and from the BTSP. The Hall B tritium cell would be filled and transported in the same manner. The storage system employed in Hall A can also be used in Hall B. This system allowed the target cell to be removed from the beam line for longer term storage (up to a few months) during accelerator down periods. It also simplifies packaging and unpackaging operations associated with shipment of the cell. 

\newpage